\documentclass[fleqn,usenatbib]{mnras}

\usepackage{newtxtext,newtxmath}

\usepackage[T1]{fontenc}

\DeclareRobustCommand{\VAN}[3]{#2}
\let\VANthebibliography\thebibliography
\def\thebibliography{\DeclareRobustCommand{\VAN}[3]{##3}\VANthebibliography}

\usepackage{graphicx}	
\usepackage{amsmath}	
\usepackage{enumitem}

\newcommand{\Omegab}{\Omega_\mathrm{b}}
\newcommand{\thetas}{\theta_\mathrm{s}}
\newcommand{\phib}{\phi_\mathrm{b}}
\newcommand{\vc}{v_\mathrm{c0}}

\title[Dynamical streams in the local stellar halo]{Dynamical streams in the local stellar halo}

\author[A. M. Dillamore et al.]{
Adam M. Dillamore,$^{1,2}$\thanks{E-mail: a.dillamore@ucl.ac.uk (AMD)}
Jason L. Sanders,$^{1}$
Vasily Belokurov,$^{2}$
and Hanyuan Zhang$^{2}$
\\
$^{1}$Department of Physics and Astronomy, University College London, London, WC1E 6BT, UK\\
$^{2}$Institute of Astronomy, University of Cambridge, Madingley Road, Cambridge CB3 0HA, UK
}

\date{Accepted XXX. Received YYY; in original form ZZZ}

\pubyear{\the\year{}}

\begin{document}
\label{firstpage}
\pagerange{\pageref{firstpage}--\pageref{lastpage}}
\maketitle

\begin{abstract}
Co-moving groups of stars (streams) are well known in the velocity space of the disc near the Sun. Many are thought to arise from resonances with the Galactic bar or spiral arms. In this work, we search for similar moving groups in the velocity space of the halo, at low angular momentum. From the asymmetry of the radial velocity distribution $v_R$, we identify two inward-moving streams with $v_R<0$ and small $|v_\phi|$. These are projections of the `chevrons' previously discovered in radial phase space $(R,v_R)$. A test particle simulation in a realistic Milky Way potential with a decelerating bar naturally produces analogues of these features, and they are observed across a wide range of metallicity. They are therefore very likely to be dynamical streams created by trapping in the bar's resonances. Specifically, they occupy regions of phase space where orbits are trapped in the corotation and outer Lindblad resonances respectively. By tracing these streams across a range of radii in $(R,v_R)$ space, we fit resonant orbits to their tracks in a flexible potential with variable bar pattern speed. This allows us to simultaneously constrain the mass profile of the Milky Way for $r\lesssim20$~kpc and the pattern speed $\Omegab$. We estimate the mass enclosed within $r=20$~kpc to be $M_{20}=(2.17\pm0.21)\times10^{11}M_\odot$, and the pattern speed to be $\Omegab=31.9_{-1.9}^{+1.8}$~km/s/kpc. Our fitted potential is in excellent agreement with previous results, while we favour a slightly slower pattern speed than most recent estimates.
\end{abstract}




\section{Introduction}\label{section:introduction}

In a totally axisymmetric galaxy in perfect equilibrium, the distribution of stars in velocity space should be symmetric in radial velocity $v_R$. It has been known for decades that this is not the case in the Milky Way. The velocity space in cylindrical coordinates $(v_R,v_\phi)$ of stars in the disc near the Sun contains multiple unbound \textit{moving groups} or \textit{streams} sharing kinematic properties with various stars and open clusters \citep{antoja2008}. These include Sirius, the Hyades \citep{eggen1958I}, Arcturus \citep{eggen1971}, and $\zeta$ Herculis \citep{eggen1958II}. The latter, usually referred to as the Hercules stream \citep{fux01}, is a group of stars with relatively low azimuthal speed ($v_\phi\sim180$~km/s) and a net \emph{outward} streaming motion (with $v_R>0$). These moving groups imply a lack of phase mixing in the local disc, and can broadly be explained by two alternative mechanisms: dissolution of open clusters \citep{eggen1965}, or dynamical interactions with the Galactic bar or spiral arms \citep[e.g.][]{dehnen1998,monari2016}. For a given star, the bar becomes dynamically important near its resonances, where its pattern speed (frequency of rotation) is commensurate with the orbital frequencies of the star (i.e. related by some integer ratio). Important resonances include the corotation (CR) and outer Lindblad resonances (OLR). Resonances can trap particles, resulting in overdensities of stars near resonant orbits \citep{binney_tremaine}. We summarise the theory behind this process in Section~\ref{section:theory}.

Astrometric measurements from the \textit{Hipparcos} space telescope \citep{hipparcos} of $\sim10^4$ nearby stars \citep{dehnen1998,skuljan1999} allowed the velocity space in the Solar neighbourhood to be mapped and modelled in more detail. Various works have associated the Hercules stream with the OLR \citep{dehnen2000,fux01}, which would imply that the bar's pattern speed is $\Omegab\approx53$~km/s/kpc \citep{dehnen2000}. Improved data from the \textit{Gaia} Observatory \citep{gaia} has spurred a renewed interest in modelling features in local velocity space to probe global properties of the Galaxy \citep{khoperskov2020,trick2021,trick2022,wheeler2022}. The majority of recent works have concluded that the Hercules stream is likely to have arisen from trapping in the CR \citep{Pe17,monari19,binney2020,chiba2021_treering,d'onghia2020}, which may require the bar's pattern speed to be slowing with time \citep{Fr19,chiba2021}. This has led to estimates for the pattern speed from Hercules being revised down to $\Omegab\approx35$~km/s/kpc \citep{binney2020,chiba2021_treering}. This is more consistent with recent values derived from direct observations of the bar, in the range $\Omegab\approx33-41$~km/s/kpc \citep{Po17,bovy2019,Sa19,Cl22,leung2023,zhang24}. A handful of new moving groups have also been identified from \textit{Hipparcos} and \textit{Gaia} data \citep{gaia_dr2_disc}. These include the inward-moving `Horn' and the high-$v_\phi$ `Hat' \citep[e.g.][]{dehnen2000,monari19,Fr19,hunt2019,laporte2020,khalil2024}. The Hat is consistent with being produced by the OLR \citep{monari19,trick2021} if the CR is responsible for the Hercules stream. Spiral arms have been shown to aid the reconstruction of the observed features in simulations, including the Sirius moving group \citep{hunt2018,hunt2019,khalil2024}. The moving groups observed locally have also been traced across different radii and azimuths \citep[e.g.][]{ramos2018,bernet2022}, which may allow the potential to be mapped over a wider area.

Substructure in velocity space at lower angular momentum was also discovered with \textit{Gaia} data. A population of stars on highly eccentric orbits dominates the stellar halo near the Sun, named \textit{Gaia} Sausage-Enceladus \citep[GSE;][]{belokurov2018,helmi2018}. In velocity space in spherical coordinates $(v_r,v_\phi)$, this population is centred at $v_\phi\sim0$~km/s and is elongated in $v_r$, stretching to $|v_r|>300$~km/s. Comparisons with simulations suggest that it is the debris of a dwarf galaxy of stellar mass up to $\sim10^{9}M_\odot$, which merged with the Milky Way $\sim8-11$~Gyr ago \citep[e.g.][]{belokurov2018,fattahi2019,belokurov2020,dillamore2022,deason2024}.

Even as early as the discovery of GSE, there were hints of finer structure within its region of velocity space. Gaussian mixture models fitted to the velocity distribution exhibit residual excesses at $v_r<0$, suggesting a bias towards inward-moving stars in the Solar neighbourhood \citep[see Fig.~3 in][]{belokurov2018}. This asymmetry is visible even down to low metallicities \citep[$\mathrm{[M/H]}<-2$;][]{zhang2023_vmp}. More recently, the third data release from \textit{Gaia} \citep[DR3;][]{gaia_dr3} included line-of-sight velocity measurements for $\sim30$ million stars by the Radial Velocity Spectrograph \citep[RVS;][]{gaia_rvs}. \citet{belokurov_chevrons} used this data to reveal a series of overdensities in the radial phase space $(r,v_r)$ of low-angular momentum stars. These `chevrons' appear most clearly at $v_r<0$, and resemble the structures which naturally result from the phase mixing of merger debris on highly eccentric orbits, such as GSE \citep[see e.g.][]{dong-paez2022}. However, debris from the GSE merger should have been undergoing phase mixing for $\gtrsim8$~Gyr, which would result in a much greater dynamical age than is observed. Specifically, \citet{donlon2023} showed that the asymmetry in $v_r$ should be erased over such a long period, calling into question whether these chevrons could have resulted from the GSE merger event. \citet{dillamore2023,dillamore2024} proposed an alternative mechanism for the formation of the chevrons, showing that they could have instead resulted from bar resonances trapping stars on highly eccentric orbits. We demonstrated using test particle simulations that a realistic bar is capable of producing chevrons with a close resemblance to the observed features. A decelerating bar helps to strengthen these features by trapping more stars \citep{chiba2021,dillamore2024}. The excess of stars at $v_r<0$ near the Sun is a natural prediction of this model, equivalent to the asymmetric moving groups in the disc such as the Hercules stream.

In this paper we investigate the asymmetric structure of the halo's velocity space in more detail, and compare the locations of moving groups in the halo to the regions where orbits are trapped in resonances. We also fit orbits to the tracks of the moving groups over a range of radii, using a flexible potential and pattern speed. This is analogous to fitting orbits to stellar streams produced by the tidal disruption of globular clusters in order to constrain the Galactic potential \citep[e.g.][]{koposov2010,malhan2019}. However, the additional requirement that the orbits be in specific resonances allows us to simultaneously constrain the pattern speed $\Omegab$ along with the Milky Way's potential in the regions explored by the orbits. This is a novel approach to inferring properties of both the Galactic bar and the dark matter halo. 

The rest of this paper is arranged as follows. In Section~\ref{section:theory} we summarise the theory of resonant trapping and calculate the regions of local velocity space in which orbits are trapped in resonances. Section~\ref{section:data} briefly describes the \textit{Gaia} data used in this paper, and Section~\ref{section:simulation} outlines the setup of the test particle simulation used for comparison. In Section~\ref{section:results} we present and discuss the results from each. We use the observations to constrain the Milky Way's potential and bar's pattern speed in Section~\ref{section:fitting}, and finally summarise our conclusions in Section~\ref{section:conclusions}.

\begin{figure*}
  \centering
  \includegraphics[width=0.74\textwidth]{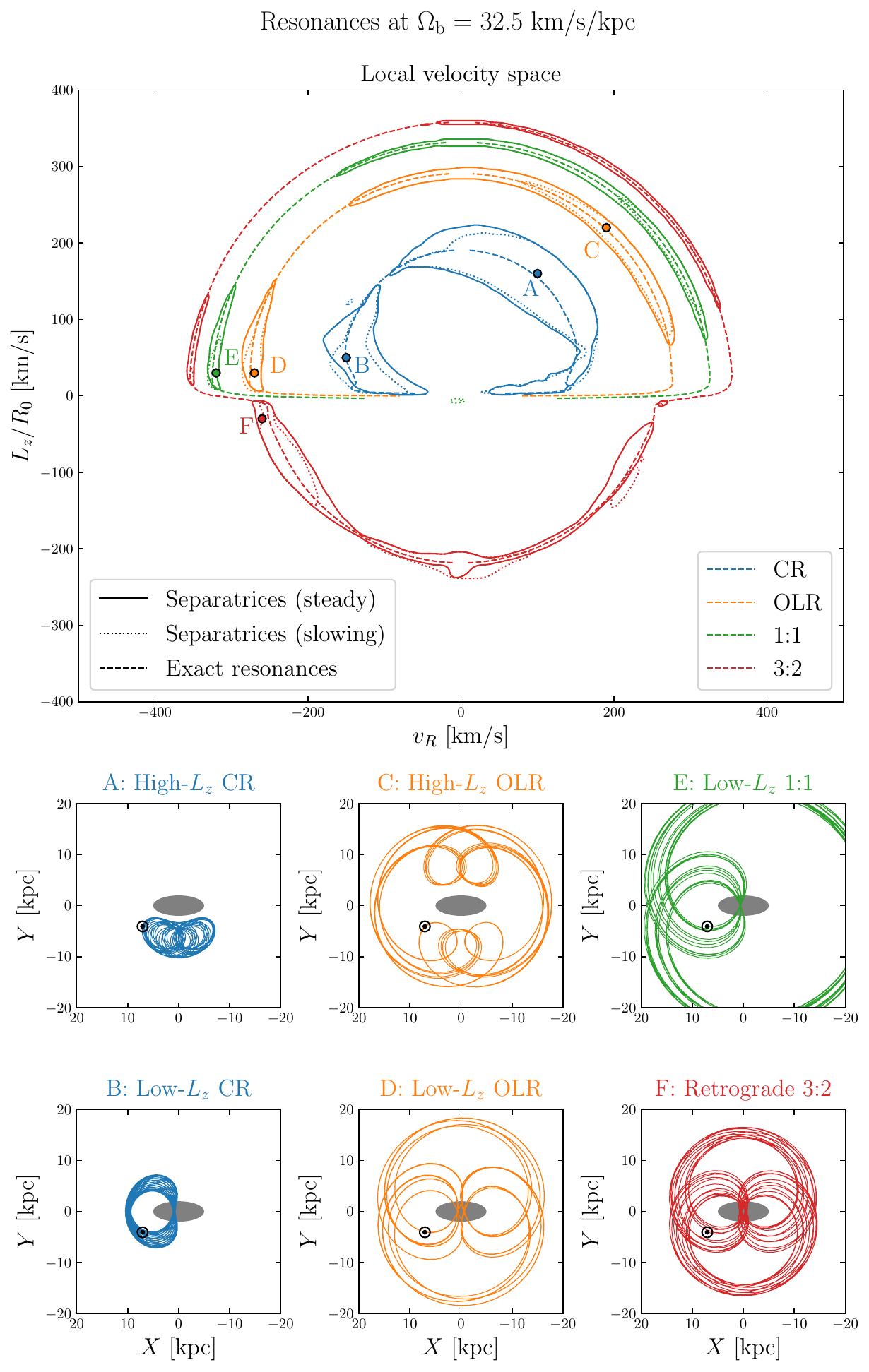}
  \caption{\textbf{Top panel:} resonances mapped in velocity space at the Sun's location. The potential is that of \citet{hunter2024}, with a steadily rotating bar at pattern speed $\Omegab=32.5$~km/s/kpc. Solid and dotted lines indicate the separatrices with a steady and slowing bar respectively. Dashed lines show the exact resonances in the limit of a weak bar. The six points marked A-F are selected as initial conditions for integrating resonant orbits for illustration. The ratios 1:1 and 3:2 correspond to the resonances $l/m=1$ and $l/m=3/2$ respectively. \textbf{Other panels:} orbits integrated from points A-F, shown in the Galactic plane in the frame corotating with the bar (clockwise in this view). The bar is marked with a grey ellipse and the Sun by the $\odot$ symbol. For prograde orbits ($L_z>0$), the resonances are split into two groups: at large $L_z$ with $v_R\gtrsim0$, and small $L_z$ with $v_R<0$. The corresponding orbits have different orientations with respect to the bar.}
   \label{fig:separatrices}
\end{figure*}

\section{Theory}\label{section:theory}
\subsection{Summary of resonant trapping}
A quasi-periodic orbit can be described by a set of three frequencies, $(\Omega_R,\Omega_\phi,\Omega_z)$, describing radial oscillations, azimuthal motion (circulation) and vertical oscillations respectively. Resonances occur when these frequencies exist in some integer ratio, sometimes involving another frequency. For resonances with a bar rotating at pattern speed $\Omegab$, the azimuthal frequency relative to the bar $\Omega_\phi-\Omegab$ is important. In this paper we consider resonances of the form
\begin{equation}\label{eq:resonances}
    l\Omega_R+m(\Omega_\phi-\Omegab)=0,
\end{equation}
where $l$ and $m$ are integers which index the resonances. Important resonances include the corotation resonance (CR; $l/m=0$), the outer Lindblad resonance (OLR; $l/m=1/2$), and the inner Lindblad resonance (OLR; $l/m=-1/2$).

The dynamics of a particle near the $(l,m)$ resonance can best be described by the \textit{slow angle},
\begin{equation}\label{eq:slow_angle}
    \thetas\equiv l\theta_R+m(\theta_\phi-\phi_\mathrm{b}).
\end{equation}
Here $\theta_R$ and $\theta_\phi$ are the angle variables (conjugate to the actions $J_R$ and $J_\phi$) describing motion in the radial and azimuthal directions (with $\dot{\theta}_R=\Omega_R=$~constant, $\dot{\theta}_\phi=\Omega_\phi=$~~constant), and $\phi_\mathrm{b}$ is the angle of the bar (such that $\dot{\phi}_\mathrm{b}=\Omegab$). Thus on the exact resonance, equation~\eqref{eq:resonances} determines that $\thetas$ is constant.

Resonant perturbation theory \citep[e.g.][]{Ly73,chirikov1979,Co97,chiba2021,hamilton2023} shows that the dynamics of the slow angle near a resonance with a quadrupole bar obey the equation of a pendulum,
\begin{align}   
    \ddot{\theta}_\mathrm{s}&\propto\mathrm{sin}\,\thetas.\label{eq:pendulum}
\end{align}
As with a pendulum, two classes of behaviour are possible: \textit{circulation} and \textit{libration}. Circulation is when $\thetas$ passes through its full range of values between 0 and $2\pi$, and occurs when $|\dot{\theta}_\mathrm{s}|$ is large. Hence this is the behaviour of a particle far from the resonance. By contrast, a librating $\thetas$ oscillates about the stable minimum of the pendulum `potential' and only explores a limited range of values. In this case, the particle is said to be \textit{trapped} by the resonance. The simplest example is the CR, for which $\thetas=2(\theta_\phi-\phib)$.\footnote{$m$ is often set to 2 for the CR to match the inner and outer Lindblad resonances \citep[e.g.][]{chiba2021}.} From equation~\eqref{eq:pendulum} it can be seen that a particle trapped in this resonance will have the azimuthal angle $(\theta_\phi-\phib)$ exploring a range $<\pi$, meaning that the orbit will be largely limited to one half of the galaxy (in the bar's frame). Given a set of orbits, we can therefore categorise them as trapped or untrapped. Trapped orbits can be seen as occupying volumes in phase space bounded by surfaces known as \textit{separatrices}. Separatrices can be visualised by fixing four of the six phase space coordinates and showing them as a function of the other two. For resonances with the bar it is most natural to fix the spatial coordinates (e.g. to the Sun's position) and the vertical velocity $v_z$ (e.g. to zero), and plot the velocity space in the Galactic plane $(v_R,v_\phi)$. Previous works have used this method to map bar resonances in the disc \citep{monari2017,binney2020,chiba2021}, but have ignored other regions of velocity space (e.g. with small $|v_R|$). We now extend maps of the bar resonance separatrices to areas of velocity space occupied by the halo.

\subsection{Calculation of separatrices}
We determine whether a given orbit is trapped by directly integrating it forward in time in a barred potential using \textsc{agama} \citep{agama}. We generate a 2D grid of orbits in velocity space $(v_R,v_\phi)$ with initial positions equal to the Sun's location and vertical velocity $v_z\approx0$~km/s. The in-plane velocity components $v_R$ and $v_\phi$ range between -400 and 400 km/s. These orbits are directly integrated forward from their initial conditions at the present day for 4 Gyr.

For each resonance $(l,m)$ we calculate the slow angle $\thetas$ as a function of time (equation~\eqref{eq:slow_angle}) using \textsc{agama}'s \textsc{ActionFinder} function. We define an orbit as trapped if $\thetas$ returns to its initial value twice without completing a circulation through $2\pi$. For a trapped orbit this usually corresponds to one complete libration. An orbit which passes through a range of $2\pi$ is classed as untrapped by the resonance in question (though it may be trapped by a different resonance). Note that for $l\neq0$ the azimuthal angle $\theta_\phi-\phib$ can circulate through $2\pi$ even if the slow angle librates, so a non-CR trapped orbit still explores all azimuths relative to the bar.

\subsection{Milky Way model}\label{section:Milky_way_model}
In this section we use the potential fitted to the Milky Way by \citet{hunter2024}. This includes the bar model fitted by \citet{sormani2020} to the made-to-measure model of \citet{Po17}. This bar consists of an X-shaped boxy-peanut bulge, a short bar, and a long bar. The potential model also features Sgr A*, a nuclear star cluster, a nuclear stellar disc, two exponential stellar disc components, a gas disc, and a dark matter halo with an Einasto profile. We set the position of the Sun to $R_0=8.178$~kpc \citep{gravity2019} and the azimuth of the Sun relative to the bar's major axis to $\phi_\odot=30^\circ$ \citep[consistent with][]{wegg2015}. We found that changing this angle by $\sim2^\circ$ had only small effects on the separatrices. The potential has a circular speed at the Sun's radius of $v_\mathrm{c}(R_0)=229$~km/s, consistent with \citet{eilers2019}. We perform the integrations with both a steadily rotating bar and a decelerating bar. In this section we set the current pattern speed of the bar to $\Omegab=32.5$~km/s/kpc, close to several recent estimates \citep[e.g.][]{Sa19,binney2020,zhang24}. We parameterise the deceleration of the slowing bar model with the dimensionless deceleration parameter, $\eta\equiv-\dot{\Omega}_\mathrm{b}/\Omegab^2$. This is set to $\eta=0.003$, consistent with \citet{chiba2021} and \citet{zhang25}. In Section~\ref{section:fitting} we will allow $v_\mathrm{c}(R_0)$ and $\Omegab$ to vary along with the dark matter halo profile.

\subsection{Resonances in local velocity space}
In the top panel of Fig.~\ref{fig:separatrices} we show resonances in cylindrical polar velocity space ($v_R$, $v_\phi$) at the Sun's position. Throughout this paper we plot the scaled angular momentum $L_z/R_0=v_\phi R/R_0$ instead of $v_\phi$ directly. While these are equal at the Sun's radius, $L_z/R_0$ is approximately conserved along a star's orbit (outside the bar) so any substructure is less blurred by observing a sample over a large volume. Each colour corresponds to a different resonance, with separatrices indicated by solid (constant $\Omegab$) and dotted (slowing $\Omegab$) lines. All orbits trapped by a given resonance are enclosed by the corresponding separatrices. The dashed lines mark the exact resonant orbits in the axisymmetrised \citet{hunter2024} potential, where equation~\eqref{eq:resonances} is satisfied exactly. These are found by calculating the orbital frequencies in the axisymmetric potential and numerically solving equation~\eqref{eq:resonances}. We also select six points inside the separatrices from which we integrate and plot orbits in the frame corotating with the bar. These are chosen to illustrate a variety of trapped resonant orbits.

In the area of velocity space occupied by the disc around $(v_R, v_\phi)\sim(0,229)$~km/s, the resonant regions are centred at $v_R>0$. The CR in particular is highly asymmetric about $v_R=0$. If the bar traps an excess of stars in this asymmetric region, it would be expected to produce an outward-moving dynamical stream, and the distribution of stars will itself be asymmetric in $v_R$. Note however that an asymmetric trapped region will not necessarily result in an asymmetric distribution. The magnitude of this asymmetry is related to the slowing of the bar, and has been used to constrain the deceleration rate \citep{chiba2021}; a decelerating bar is able to trap an excess of stars in its CR and therefore results in increased asymmetry. An observed example is the Hercules Stream, a moving group with $v_R>0$ which has been associated with the CR \citep{Pe17,d'onghia2020,binney2020,chiba2021}. In this case it would be made up of orbits similar to that labelled `A'. The regions of higher resonances ($l/m>0$) are similarly centred at $v_R>0$. In each case the orbits (e.g. A and C) have pericentres aligned roughly with the minor axis of the bar.

However, each resonance also has a distinct region of trapping at $L_z/R_0\lesssim150$~km/s and $v_R<0$. These orbits (B, D and E) instead have pericentres aligned with the major axis of the bar, resulting in the orbits passing through the Solar neighbourhood inwardly. In \citet{dillamore2024} we predicted this behaviour using resonant perturbation theory in the isochrone potential, and demonstrated that trapped orbits with low-$L_z$ are expected to have $v_R<0$ near the Sun. The same is true for low-$|L_z|$ retrograde orbits (e.g. orbit F). As explained in \citet{dillamore2024}, the prograde $l/m$ resonance is continuous at $L_z=0$ with the retrograde $l/m+1$ resonance. This is because as $L_z$ changes infinitesimally from positive to negative, the orbit makes an extra circulation about the origin per pericentric passage (in the corotating frame). Hence the prograde OLR is continuous with the retrograde 3:2 resonance, and the behaviour of these respective orbits either side of $L_z=0$ is similar. Note however that with this bar potential there is no retrograde 1:1 resonance at the Sun's location to match the prograde CR.

The dotted contours indicate that the separatrices have some dependence on the deceleration rate of the bar. With $\eta=0.003$ the separatrices enclose a smaller volume of velocity space, so any groups of trapped stars would be expected to be more concentrated. This is consistent with the findings of \citet{chiba2021}.

From Fig.~\ref{fig:separatrices} we may therefore predict the presence and locations of dynamical moving groups with low $|L_z|$ and $v_R<0$, produced by the trapping of halo stars in resonances. We now proceed to search for such moving groups in data and simulations.

\section{Data}\label{section:data}
We use the data from Data Release 3 (DR3) of the \textit{Gaia} observatory \citep{gaia,gaia_dr3} including line-of-sight velocity measurements from the Radial Velocity Spectrograph \citep[RVS;][]{gaia_rvs} and distances from \citet{bailer-jones2021}. Following \citet{belokurov_chevrons} and \citet{dillamore2023}, we limit our sample to stars with parallax signal-to-noise values $\varpi/\sigma_\varpi>10$, and further than 1.5$^\circ$ from globular clusters within 5~kpc of the Sun. We also use metallicity values [M/H] for red giant stars calculated from \textit{Gaia} BP/RP spectra by \citet{andrae2023}.

We transform the 6D position and velocity measurements into a Galactocentric coordinate system. We assume that the Sun sits in the Galactic plane at a radius of $R_0=8.178$~kpc \citep{gravity2019} at an angle of $30^\circ$ to the bar \citep{wegg2015}, and moves with velocity $(v_R,v_\phi,v_z)=(-11.1, 247.4, 7.25)$~km/s. The radial and vertical components are measurements of the Sun's motion relative to the local standard of rest by \citet{schonrich2010}, and the azimuthal component is derived from distance and proper motion measurements of Sagittarius A* by \citet{gravity2019} and \citet{reid2004}.

\begin{figure}
  \centering
  \includegraphics[width=0.9\columnwidth]{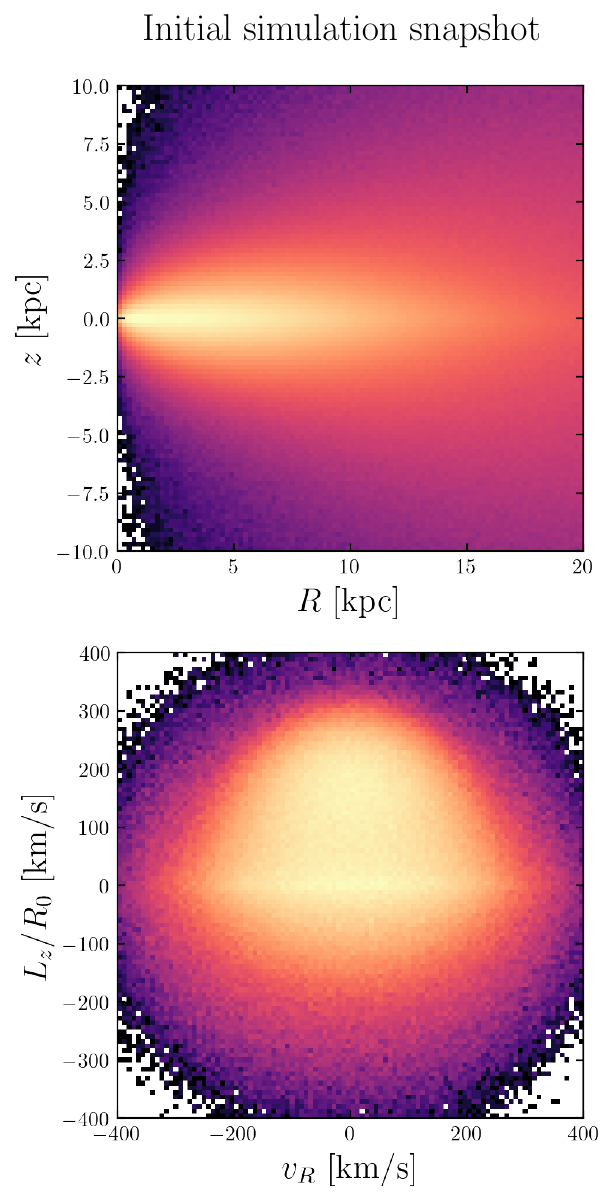}
  \caption{\textbf{Top panel:} Height $z$ vs cylindrical radius $R$ of the initial simulation distribution. The distribution is axisymmetric, so this is independent of $\phi$. \textbf{Bottom panel:} Initial velocity distribution of the simulation in the Solar neighbourhood ($|R-R_0|<1$~kpc, $|z|<2$~kpc).  Since this is an equilibrium axisymmetric distribution, it is symmetric in $v_R$.}
   \label{fig:sim_initial}
\end{figure}

\section{Simulation}\label{section:simulation}
For comparison with the data we run a test particle simulation in a potential with a smoothly growing and decelerating bar. The simulation setup is similar to that used by \citet{dillamore2024}, and we describe it briefly below. In contrast to Section~\ref{section:theory}, we initialise the particles from a steady state distribution function in an axisymmetric potential, before smoothly introducing a bar. The distribution has a range of vertical actions, so particles are no longer constrained to idealised planar orbits.

The potential consists of the axisymmetrised (azimuthally averaged) \citet{hunter2024} potential plus a time-dependent bar perturbation,
\begin{equation}
    \Phi(R,\phi,z,t)=\Phi_\mathrm{axi}(R,z)+ \Phi_\mathrm{bar}(R,\phi,z,t).
\end{equation}
The bar component $\Phi_\mathrm{bar}$ consists of the quadrupole and higher-order Fourier harmonics in $\phi$ of the \citet{hunter2024} barred potential, but with time-dependent amplitude and length. Since the size and shape of the separatrices depends on the bar properties \citep{chiba2021}, we choose this realistic fitted bar model in order to provide the best comparison to the observations.

The relative amplitude of $\Phi_\mathrm{bar}$ is smoothly increased from 0 to 1 between times $t=0$ and $t_1=2$~s\,kpc/km~$\approx2$~Gyr, following equation~4 in \citet{dehnen2000}. The pattern speed $\Omegab(t)$ then begins to smoothly decrease according to the prescription used by \citet{dillamore2024}, briefly described below. The pattern speed begins to decelerate between times $t_1\approx2$~Gyr and $t_2\approx3$~Gyr, following equation (C3) in \citet{dillamore2024}. At $t>t_2$ we keep the dimensionless deceleration parameter $\eta$ constant. In a potential with a flat rotation curve, a constant $\eta$ corresponds to the corotation radius moving outwards at a constant rate. As in Section~\ref{section:Milky_way_model} we set $\eta=0.003$, resulting in $\Omegab$ decreasing from 60~km/s/kpc to $\approx30$~km/s/kpc between time $t_1\approx2$~Gyr and the end of the simulation at $t_\mathrm{f}\approx8$~Gyr.

As the bar slows, we also adjust the physical scale of the bar to account for the increasing corotation radius $R_\mathrm{CR}$. We simply adjust the relative scale of the bar according to $S_\mathrm{bar}(t)=\Omega_0/\Omegab(t)$, where $\Omega_0$ is the pattern speed of the fiducial `present-day' snapshot. Hence when $\Omegab=\Omega_0$, our potential matches that of \citet{hunter2024}. If the rotation curve was exactly flat, this would keep the ratio $R_\mathrm{CR}/S_\mathrm{bar}$ constant. As in Section~\ref{section:theory} we set the fiducial pattern speed to $\Omegab=32.5$~km/s/kpc, corresponding to a time of $t\approx7.2$~Gyr after the onset of bar growth in the simulation. This is roughly consistent with the age of the bar inferred by \citet{sanders2024}.

The initial distribution of particles is generated from a steady-state distribution function (DF) $f$ in the axisymmetric potential consisting of two components,
\begin{equation}
    f_\mathrm{tot}=f_\mathrm{disc}+f_\mathrm{halo}.
\end{equation}
In this paper we focus on highly eccentric orbits in the Solar neighbourhood. We therefore choose the parameters of the two DF components to provide a large number of stars on these orbits, instead of aiming to accurately model the distribution of stars in the Milky Way.

The disc component $f_\mathrm{disc}$ is an action-based \textsc{Exponential} DF implemented in \textsc{agama} \citep{agama},
\begin{align}
    f_\mathrm{disc}&\propto\tilde{J}^3\mathrm{exp}\left(-\frac{\tilde{J}}{J_{\phi,0}}\right)\mathrm{exp}\left(-\frac{\tilde{J}K(\mathbf{J})}{J_{r,0}^2}\right)\mathrm{exp}\left(-\frac{\tilde{J}J_z}{J_{z,0}^2}\right),\\
    \tilde{J}&\equiv|J_\phi|+J_r+0.25J_z,\\
    K(\mathbf{J})&\equiv\begin{cases}
        J_r & J_\phi\geq0\\
        J_r-J_\phi & J_\phi<0.
    \end{cases}
\end{align}
We set the scale-length to $J_{\phi,0}=500$~kpc~km/s, the vertical action scale to $J_{z,0}=200$~kpc~km/s, and the radial action scale to $J_{r,0}=800$~kpc~km/s. These correspond to a radial scale of $R_\mathrm{disk}\approx2.2$~kpc, and vertical and radial velocity dispersions in the Solar neighbourhood of $\sigma_z\sim10$~km/s and $\sigma_r\sim100$~km/s respectively. The large value of $J_{r,0}$ ensures that we generate a large number of stars on eccentric orbits.

The halo component $f_\mathrm{halo}$ is a \textsc{DoublePowerLaw} DF,
\begin{align}
    f_\mathrm{halo}(\textbf{J})&=\frac{M}{(2\pi J_0)^3}\left[1+\left(\frac{J_0}{h(\textbf{J})}\right)\right]^\Gamma \left[1+\left(\frac{h(\textbf{J})}{J_0}\right)\right]^{-B},\\
    h(\textbf{J})&\equiv h_rJ_r+h_zJ_z+(3-h_r-h_z)|J_\phi|.
\end{align}
The inner and outer slopes are $\Gamma=1$ and $B=4$, and the characteristic action is set to $J_0=500$~kpc~km/s. We set the anisotropy and flattening coefficients to $h_r=0.2$ and $h_z=2.5$. The small value of $h_r$ gives a highly radially anisotropic distribution, including many stars on GSE-like orbits. The choice of $h_z$ gives a highly flattened distribution. We set the mass normalisation $M$ equal to the total mass of the disc component. In Fig.~\ref{fig:sim_initial} we show the initial spatial distribution (top panel) and velocity distribution of stars within the region $|R-R_0|<1$~kpc, $|z|<2$~kpc (bottom panel). Since the initial distribution is axisymmetric, this distribution is independent of $\phi$.

We integrate the orbits of the particles between times $t=0$ and $t=t_\mathrm{f}$ in the decelerating barred potential $\Phi$ using \textsc{agama}. We take a fiducial snapshot at $t\approx7.2$~Gyr, at which the pattern speed is $\Omegab=32.5$~km/s/kpc. The results are shown along with the data in Section~\ref{section:results}.

\begin{figure*}
  \centering
  \includegraphics[width=\textwidth]{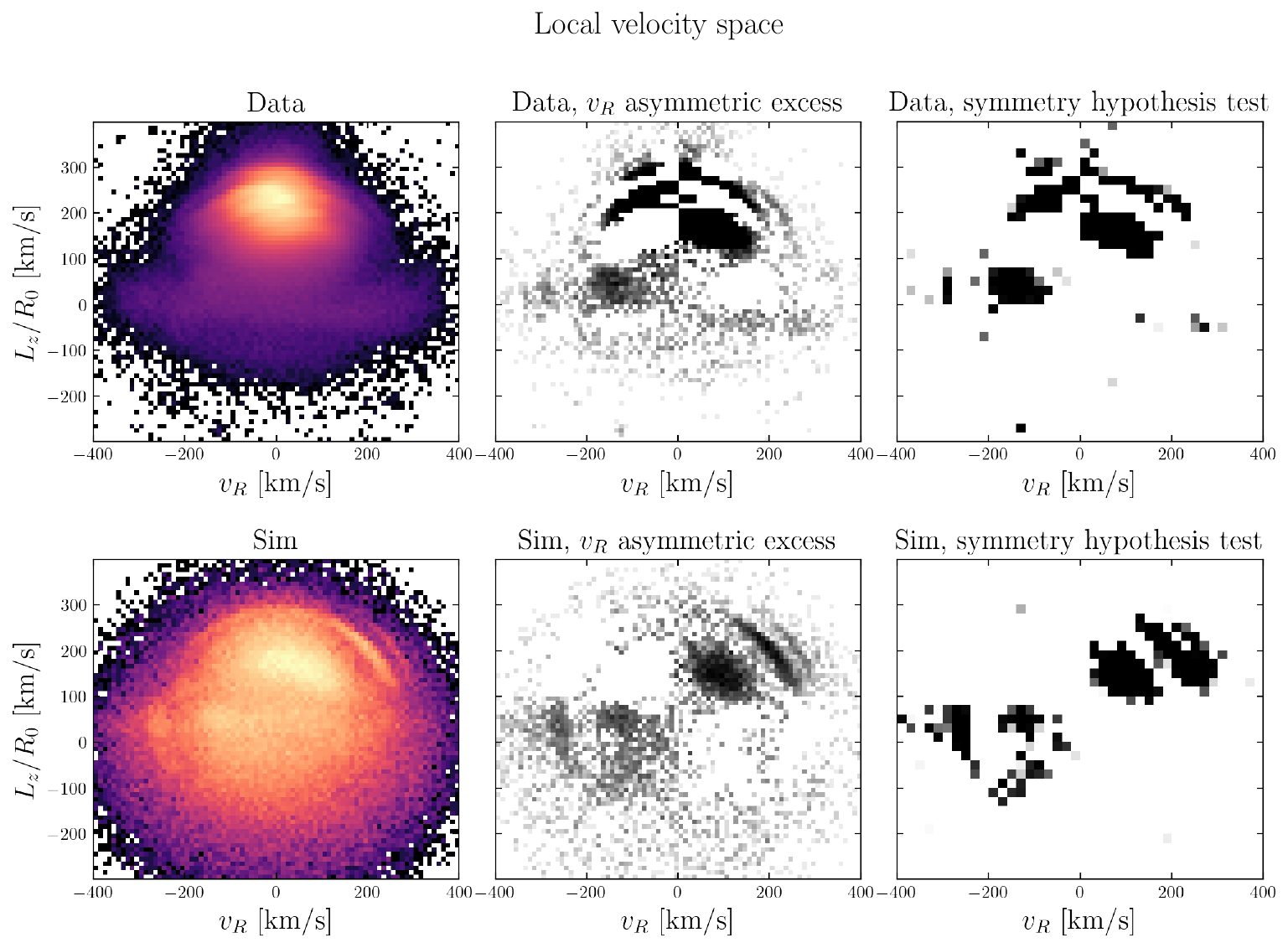}
  \caption{\textbf{Left-hand column:} velocity space in the Solar neighbourhood of the \textit{Gaia} data (top panel) and simulation in its fiducial snapshot (bottom panel). The data contains a series of clumps and ridges, some of which are reproduced in the simulation. \textbf{Middle column:} The antisymmetric component of the histograms $n_\mathrm{antisym}(v_R,v_\phi)\equiv n(v_R, v_\phi)-n(-v_R, v_\phi)$. Only positive values of $n_\mathrm{antisym}$ (i.e. where there is an excess) are shown by grey/black pixels. \textbf{Right-hand column:} $p$-values of a hypothesis test in each pixel for symmetry between opposite values of $v_R$. The null hypothesis is that the sign of $v_R$ for each star is drawn from a binomial distribution with $p$ no greater than 0.5. Grey/black pixels indicate where the $p$-value is less than 0.01.} 
   \label{fig:data_sim}
\end{figure*}

\begin{figure*}
  \centering
  \includegraphics[width=\textwidth]{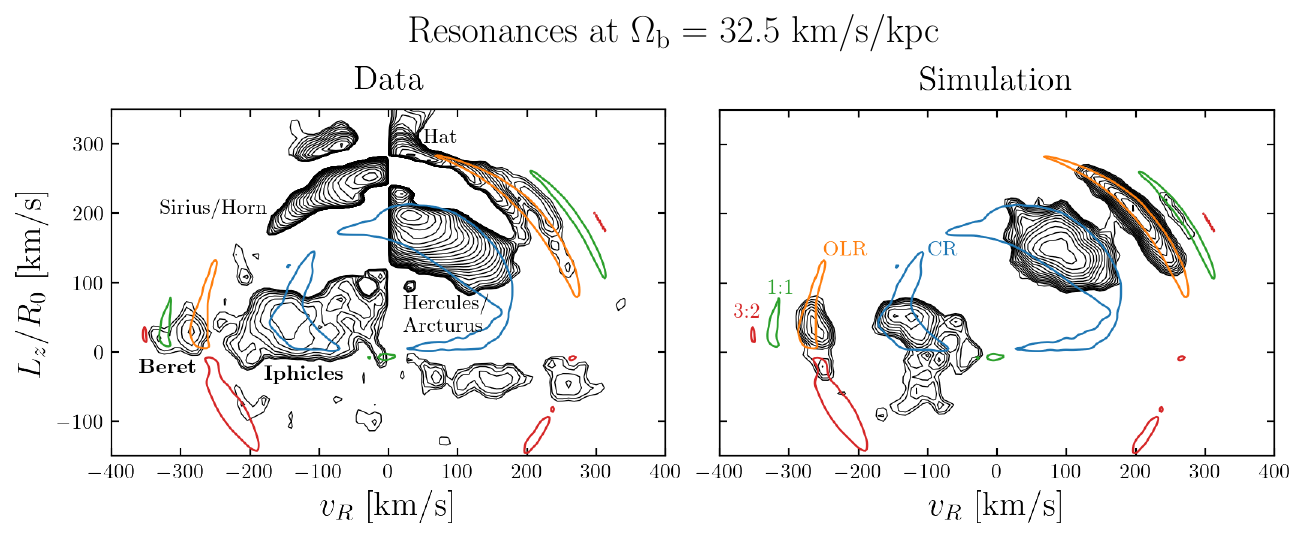}
  \caption{\textbf{Left-hand panel:} antisymmetric component of the $(v_R, L_z/R_0)$ distribution for observed stars in the Solar neighbourhood. Moving groups are labelled, including the two named in this work, Iphicles and the Beret. \textbf{Right-hand panel:} the same for the simulation, with the separatrices labelled by their resonances. In both panels the coloured curves are the resonant separatrices. To match the simulation these are calculated with a decelerating bar ($\eta=0.003$) so enclose smaller volumes than in the steadily rotating case.} 
   \label{fig:data_sim_seps}
\end{figure*}

\begin{figure*}
  \centering
  \includegraphics[width=\textwidth]{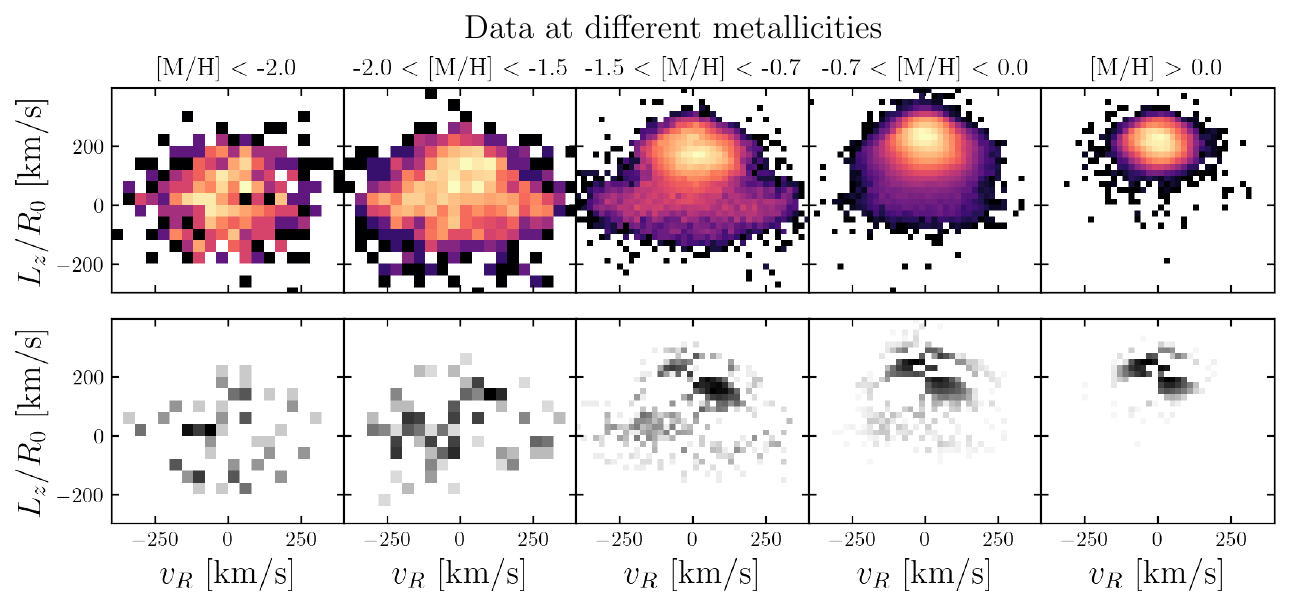}
  \caption{\textbf{Top row:} local velocity space of red giant stars in different bins of metallicity [M/H], as calculated by \citet{andrae2023}. \textbf{Bottom row:} the antisymmetric components $n_\mathrm{antisym}$ of the above distributions. The low-$L_z$ antisymmetric features are seen across a wide range of metallicities: Iphicles at $(v_R,L_z/R_0)\approx(-150,50)$~km/s is clearly visible at all metallicities below [M/H]~$=0$, and the Beret at $\approx(-300,25)$~km/s is discernible in all bins below [M/H]~$=-0.7$.}
   \label{fig:data_metallicity}
\end{figure*}

\begin{figure*}
  \centering
  \includegraphics[width=\textwidth]{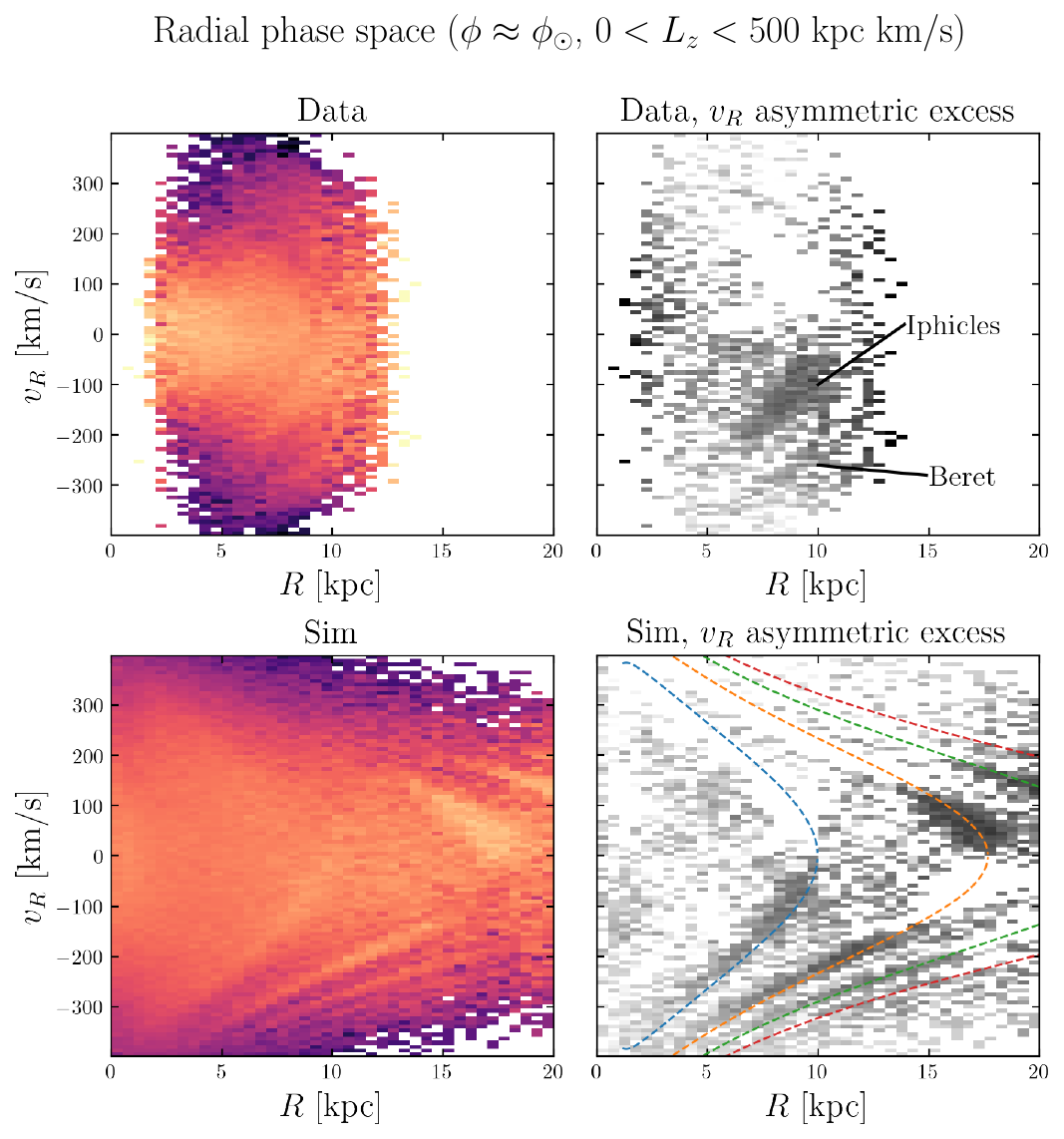}
  \caption{\textbf{Left-hand column:} radial phase space $(R,v_R)$ of data (top panel) and the simulation (bottom panel), for low-angular momentum prograde stars ($0<L_z<500$~kpc~km/s). We apply the additional cuts $|\phi-\phi_\odot|<\pi/12$ and $|z|<2$~kpc to limit the samples to the Solar neighbourhood azimuthally and vertically. \textbf{Right-hand column:} the antisymmetric components of the same histograms in $v_R$, as in Fig.~\ref{fig:data_sim}. The lower panel shows the tracks of the exact resonant orbits for $L_z=250$~kpc~km/s with dashed lines, where the colours correspond to the same resonances as in Fig.~\ref{fig:separatrices}. In the simulation, clear `chevron' structures can be seen predominantly at $v_R<0$. These align exactly with the principal resonances. Similar structures are also seen in the data, as first identified by \citet{belokurov_chevrons}. These are the extensions of `Iphicles' and the 'Beret' across different radii.} 
   \label{fig:R_vR}
\end{figure*}

\begin{figure*}
  \centering
  \includegraphics[width=\textwidth]{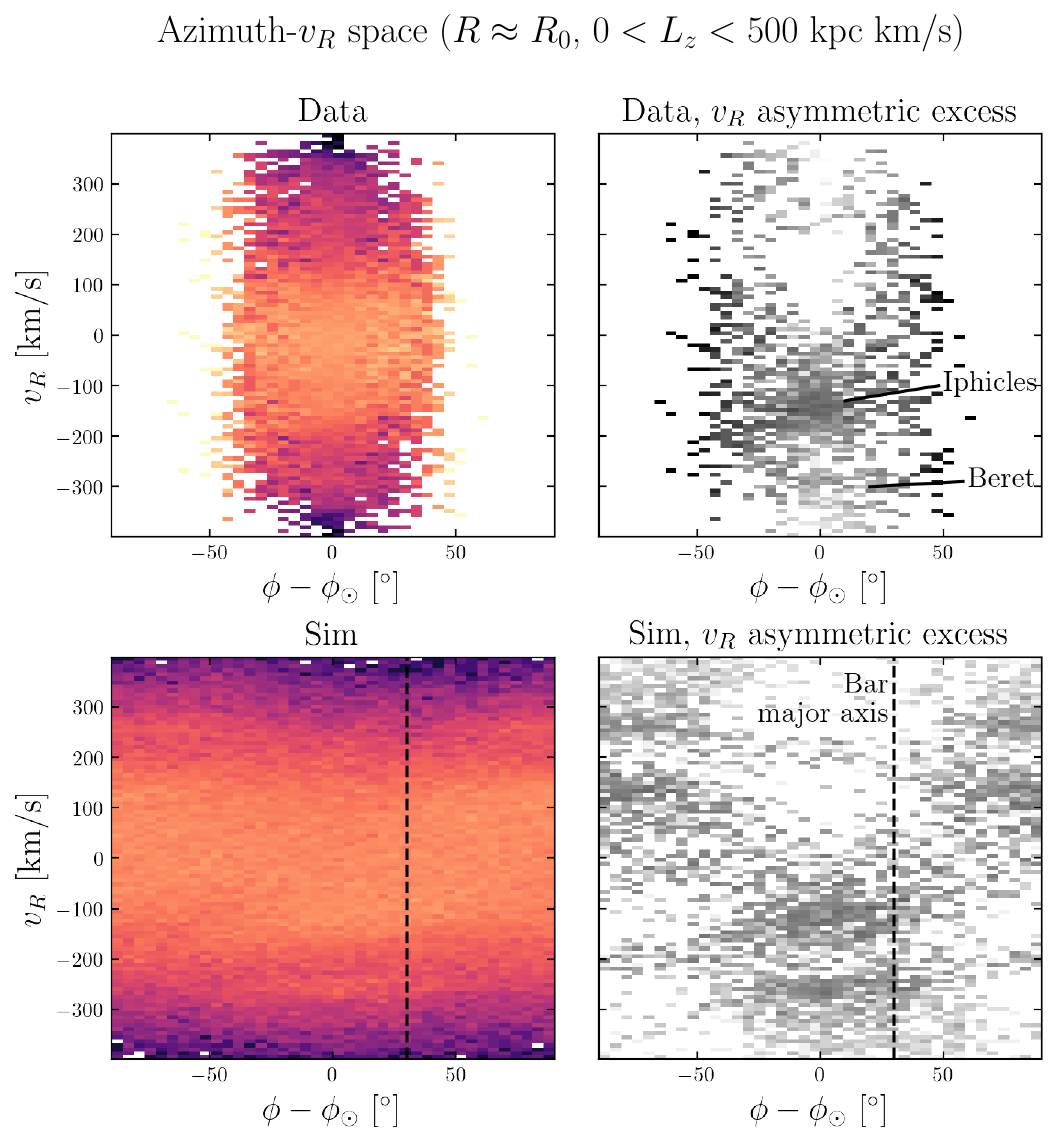}
  \caption{Like Fig.~\ref{fig:R_vR}, but showing $v_R$ vs Galactocentric azimuth relative to the Sun, $\phi-\phi_\odot$. Stars are selected from the radial range $|R-R_0|<1$~kpc and the vertical range $|z|<2$~kpc. The vertical dashed black line in the lower panels indicates the major axis of the bar in the simulation. In the simulation the CR and OLR features at $v_R<0$ can be traced $\sim40^\circ$ in each direction, beyond which they are visible at $v_R>0$. The azimuth of this flip is slightly offset from the bar's major axis, likely because of its deceleration. In the data the Iphicles and Beret features can be seen across a few tens of degrees, beyond which the data becomes too sparse.} 
   \label{fig:phi_vR}
\end{figure*}

\begin{figure*}
  \centering
  \includegraphics[width=\textwidth]{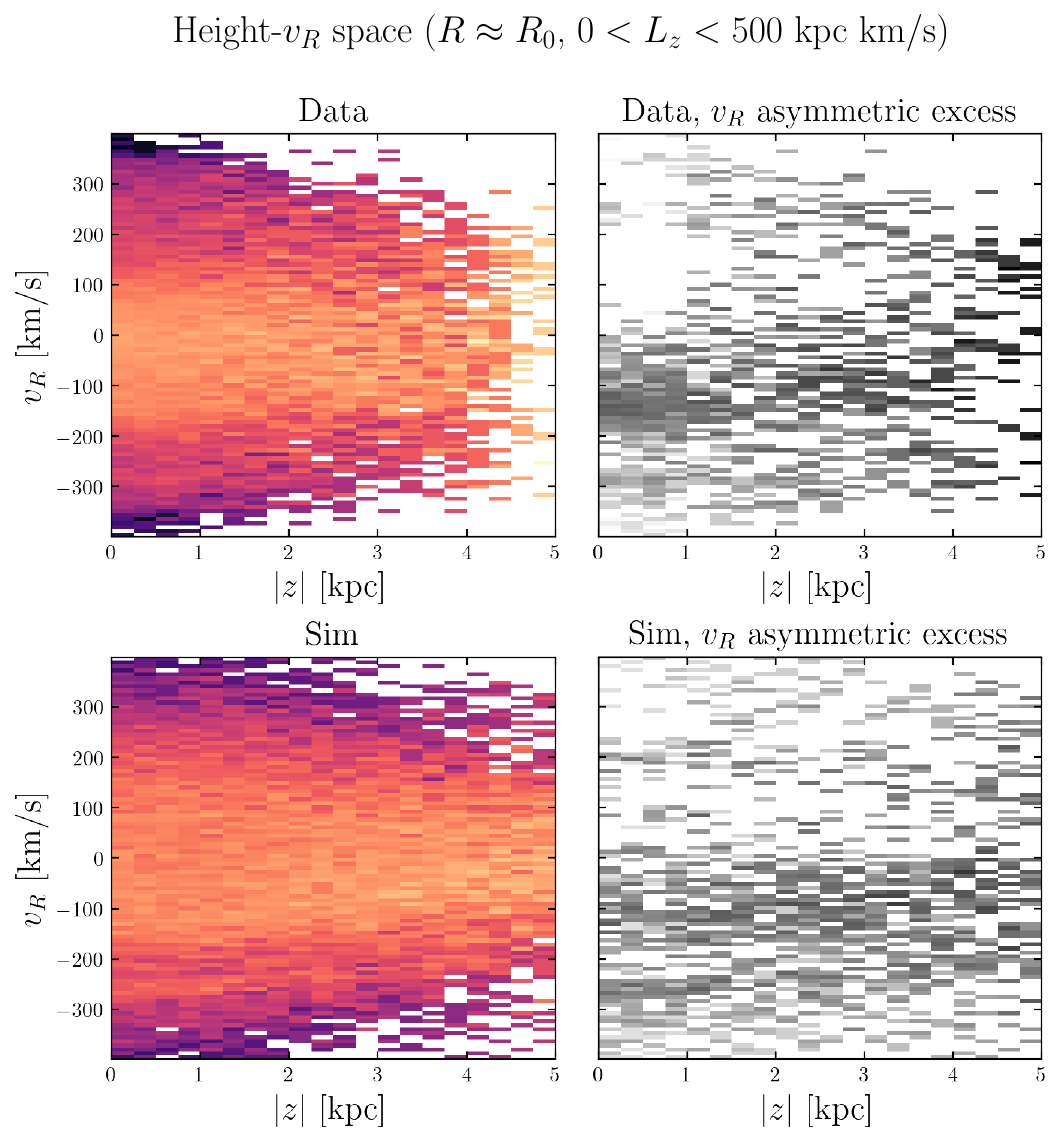}
  \caption{Like Figs.~\ref{fig:R_vR} and \ref{fig:phi_vR}, but showing $v_R$ vs distance from the Galactic plane, $|z|$. Stars are selected from the radial range $|R-R_0|<1$~kpc and the azimuthal range range $|\phi-\phi_\odot|<\pi/12$. The Iphicles and Beret features in the data are visible as asymmetric excesses at $v_R\approx-150$~km/s and $v_R\approx-300$~km/s respectively. The Iphicles feature extends at least as far as $|z|=4$~kpc. The Beret is only clearly visible up to $|z|\sim1$~kpc, with the density of stars in the sample becoming too low beyond that. The CR and OLR features at $v_R<0$ in the simulation extend well above the Galactic plane, even as far as $|z|\sim5$~kpc.} 
   \label{fig:z_vR}
\end{figure*}

\section{Results}\label{section:results}

\subsection{Local velocity space}
As demonstrated in Section~\ref{section:theory}, trapped moving groups can be identified by their asymmetry in $v_R$. While an equilibrium population of stars in an approximately axisymmetric potential would have equal numbers of stars at $\pm v_R$, a group of trapped stars will generally be unequally split between positive and negative values. We therefore focus on the \textit{antisymmetric} component of $(v_R,L_z/R_0)$ histograms, defined as $n_\mathrm{antisym}(v_R,L_z/R_0)\equiv n(v_R, L_z/R_0)-n(-v_R, L_z/R_0)$. A cell with a significantly non-zero $n_\mathrm{antisym}$ implies a lack of phase-mixing, equilibrium, and/or axisymmetry, such as trapping in resonances.

In Fig.~\ref{fig:data_sim} we show the local velocity space in the Solar neighbourhood for the data (top row) and the simulation (bottom row) at a pattern speed of $\Omegab=32.5$~km/s/kpc. The spatial cuts applied are $|R-R_0|<1$~kpc, $|\phi-\phi_\odot|<\pi/12$ and $|z|<2$~kpc. This is a volume of approximate dimensions $1\times2\times2$~kpc. No metallicity cuts are applied here. The left-hand panels show the density of stars in velocity space. Most coarsely, the data can be divided into the disc ($L_z/R_0\sim200$~km/s) and the halo, dominated by \textit{Gaia} Sausage-Enceladus (GSE; at $L_z/R_0=0$, $|v_R|\lesssim400$~km/s). Further substructure can be seen within these populations. Several clumps and ridges are seen in the disc, most of which are not symmetric about $v_R=0$. However, the GSE component also shows some asymmetric substructure.

The $v_R$ asymmetry is revealed more clearly in the middle column, where we show the antisymmetric components of the histograms $n_\mathrm{antisym}$. Only positive values (i.e. asymmetric excesses) are shown with grey/black pixels. At high $L_z/R_0$, the asymmetries lie along several arcs at approximately constant energy (i.e. $|\textbf{v}|$). These features are well-known, and include the Hercules, Sirius and Hyades streams \citep{eggen1958I,eggen1958II,dehnen2000,fux01,antoja2008}, the `Horn' and the `Hat' (see Fig.~\ref{fig:data_sim_seps}). Several of these have been associated with bar resonances \citep[see Section~\ref{section:introduction};][]{dehnen2000,Pe17,d'onghia2020,chiba2021_treering,chiba2021,monari19,trick2021} or spiral arms \citep[e.g.][]{hunt2019,khalil2024}. In addition to these features, there are also several asymmetries at $L_z/R_0<100$~km/s. Most clearly, there is a broad peak at $\approx(-150,50)$~km/s and a narrower one at $\approx(-300,25)$~km/s. The significance of these features is assessed in the right-hand panel, where we perform hypothesis tests on the asymmetry. We count the total number of stars in each pair of bins at $(\pm v_R,L_z/R_0)$, i.e. $n_\mathrm{sym}(v_R,L_z/R_0)\equiv n(v_R, L_z/R_0)+n(-v_R, L_z/R_0)$. The null hypothesis is that the sign of $v_R$ of each star in this pair of bins is binomially distributed with $p$ no greater than 0.5. The null hypothesis is therefore rejected if the count in a bin at $v_R$ is significantly larger than in the corresponding bin at $-v_R$, implying significant asymmetry. We use bins of width 20~km/s, which is larger than the typical velocity uncertainty of $\sim1-10$~km/s \citep{gaia_rvs}. Pixels where this is rejected at a 1\% level of significance are shown in grey/black. The clumps of contiguous dark pixels show where there are robust asymmetric features, including those described above. In particular, the features at $v_R<0$ and $L_z>0$ are more statistically significant than those at $v_R>0$ and $L_z<0$.

The bottom row shows the corresponding results from the simulation. Asymmetric ridges and clumps are also seen here. Most clearly, three at $L_z/R_0\gtrsim100$~km/s are at $v_R>0$, while two with lower angular momentum have $v_R<0$. Most of these features have counterparts at similar locations in the data, including the two low-$L_z$ inward-moving peaks and the Hercules Stream at $(100,150)$~km/s. We can therefore reproduce much of the observed asymmetry with bar resonances. Note however that not all features are reproduced by the simulation; most notably, the observed peaks at $L_z/R_0\gtrsim200$~km/s with $v_R<0$ are not seen in the simulation. These observed features may be explained by interactions with spiral arms \citep[e.g.][]{hunt2019,khalil2024}.

To link asymmetric features with individual resonances, we show the antisymmetric components of the velocity distributions again in Fig.~\ref{fig:data_sim_seps}. Over these we plot the separatrices of the resonances as described in Section~\ref{section:theory}, labelled in the bottom panel. We use the \citet{hunter2024} potential (not including the spiral arms) at a pattern speed of $\Omegab=32.5$~km/s/kpc. We include deceleration when calculating the separatrices for this figure, so the resonant regions match those enclosed by the dotted contours in Fig.~\ref{fig:separatrices}.

As expected, the asymmetric peaks in the simulation closely align with theoretically predicted regions of trapped orbits. The four strongest peaks correspond to the inward and outward moving (i.e. low and high $L_z$) CR and OLR families of orbits. There are also weaker peaks enclosed by the outward-moving 1:1 resonance (at $(v_R,L_z/R_0)\approx(270,200)$~km/s) and the retrograde 3:2 resonance (at $(v_R,L_z/R_0)\approx(-250,-20)$~km/s). The one region with strong asymmetry and no resonances is around $(-100,-50)$~km/s, though this is immediately adjacent to the CR region.

In the left-hand panel we label the approximate locations of the named moving groups in the disc. These include the outward-moving Hercules and Arcturus streams, and the inward-moving `Horn' and Sirius moving groups. The higher-$L_z$ `Hat' is seen at both positive and negative $v_R$. Other moving groups such as the Hyades, Pleiades and Coma Berenices are situated closer to $v_R=0$ \citep{antoja2008} so do not show clearly in the antisymmetric distribution. We refer to the low-$L_z$ moving groups at $\approx(-150,50)$~km/s and $\approx(-300,25)$~km/s as `Iphicles' and the `Beret' respectively.\footnote{Iphicles was the brother of Hercules, and the `Iphicles' moving group is associated with the Hercules Stream. The `Beret' is at a similar energy to the `Hat', but sits on the side of the velocity distribution.} These are labelled in bold.

As in the simulation, most of the asymmetries in the data can be associated with specific resonances. At this slow pattern speed ($\Omegab=32.5$~km/s/kpc), the Hercules Stream coincides with the CR, in agreement with various recent works which considered $\Omegab<40$~km/s/kpc \citep{Pe17,binney2020,chiba2021_treering}. This is also in agreement with the long slow bar model (the `Hat model') of \citet{trick2021}, for which the Hat feature is produced by the OLR. Fig.~\ref{fig:data_sim_seps} shows that this interpretation is consistent with our fiducial pattern speed, since the OLR separatrices coincide with the Hat. The story is similar at low $L_z$. At this $\Omegab$, the peak of Iphicles lies within the inward-moving CR. Likewise, the Beret peaks on the edge of the inward-moving OLR, although it is immediately adjacent to a smaller peak near the 1:1 resonance. Both Iphicles and the Beret have analogues in the simulation.

However, there are several asymmetric features in the data which are not reproduced by the simulation. Most notably, the inward-moving features in the disc such as the Sirius moving group, the Horn and part of the Hat do not have analogues in the simulation. These may be explained by spiral structure in the disc, which have been shown to produce moving groups consistent with \textit{Gaia} data \citep{hunt2019,khalil2024}. There are also two retrograde peaks not seen in the simulation, at $(150,-40)$~km/s and $(250,-50)$~km/s. The latter lies close to the retrograde 3:2 resonance, so may be created by this. The lack of such a feature in the simulation could result from the deceleration of the bar, which significantly shrinks this resonant region. The peak at $(150,-40)$~km/s does not overlap with any principal resonances at this pattern speed. Its presence is therefore difficult to explain by bar resonances alone. Note however that the hypothesis test demonstrates that these peaks are less significant than Iphicles or the Beret (see top-right panel of Fig.~\ref{fig:data_sim}); the null hypothesis of symmetry is not rejected in these areas.

A way to distinguish dynamical (e.g. resonant) streams from those produced by the disruption of clusters is by measuring the metallicity of their constituent stars. Stars from tidally stripped clusters should have similar metallicities, so a wide range of metallicities in a moving group would provide evidence of a dynamical origin. Note that the converse is not true: a lack of stars at some metallicities in a stream does not mean that a dynamical origin can be ruled out, since only certain initial orbits can be trapped into a given stream. For example, stars at high metallicity in the disc are not expected to be easily trapped into halo-like resonant orbits.

In Fig.~\ref{fig:data_metallicity} we show the $(v_R,L_z/R_0)$ distributions of observed stars from \textit{Gaia} in five bins of metallicity [M/H]. This figure is produced using the data from Table 2 of \citet{andrae2023}, which is a vetted sample of 17.5 million red giants whose metallicity measurements are precise. We use the same volume and parallax signal-to-noise cuts as in Fig.~\ref{fig:data_sim}. The top row of Fig.~\ref{fig:data_metallicity} shows the density of stars, and the bottom row shows the antisymmetric components of the distributions $n_\mathrm{antisym}$.

The presence of the asymmetries is largely a function of where stars are located in velocity space. At high metallicity ([M/H]~$>0$), the only significant stellar population is the disc, with very few stars at $L_z/R_0<100$~km/s. Correspondingly, only the asymmetries in the disc (Hercules, the Horn, the Hat etc.) are visible in the lower panel. At [M/H]~$<0$, the halo also contributes to the stellar density, with some stars around $L_z\approx0$. In the $-0.7<\mathrm{[M/H]}<0.0$ range, the Iphicles asymmetry is visible as well as the disc moving groups. While the Beret cannot be seen, there are few stars in this distribution at $|v_R|\gtrsim300$~km/s, so this is unsurprising. At $-2.0<\mathrm{[M/H]}<-0.7$, the disc moving groups begin to fade, though the Hercules Stream remains the most prominent feature. At low $L_z$, both Iphicles and the Beret can be seen in the asymmetry. This is the range of [M/H] at which GSE is most dominant \citep{naidu2020,feuillet2021}, so it is where a large proportion of the low-$L_z$, high $|v_R|$ stars are situated. It is therefore natural that this is where the low-$L_z$ asymmetries are most prominent. However, the asymmetries can be seen at even lower metallicity. At [M/H]~$<-2$, while Hercules is still visible, the most prominent asymmetry is Iphicles. The Beret is also still faintly discernible at $v_R\approx-300$~km/s. These features were previously described by \citet{zhang2023_vmp}, who showed using Gaussian mixture models that the velocity distribution at low metallicity includes a highly radial component with $v_R<0$.

In summary, all of the asymmetric features labelled in Fig.~\ref{fig:data_sim_seps} are seen across a range of metallicities, in at least three different [M/H] bins in Fig.~\ref{fig:data_metallicity}. The well-known disc moving groups dominate at high metallicity, while the low-$L_z$ inward-moving streams are most prominent at low metallicities, particularly [M/H]~$<-1$. This is a reflection of the fact that the disc and halo have higher and lower metallicities respectively.

\subsection{Spatial dependence}
We have so far focused on the slice of phase space at the location of the Sun. We now turn to investigate how the asymmetric moving groups vary as a function of spatial position in the Galaxy. \citet{bernet2022} comprehensively performed this analysis for moving groups in the disc, so we restrict ourselves to stars with low $L_z$, in the range $0<L_z<500$~kpc~km/s (i.e. $0<L_z/R_0\lesssim61$~km/s). This excludes the outward-moving Hercules/Arcturus streams, but includes Iphicles and the Beret. In Figs.~\ref{fig:R_vR}, \ref{fig:phi_vR} and \ref{fig:z_vR} we show the distributions of stars as functions of the cylindrical coordinates $R$, $\phi$ and $|z|$ respectively. In each case we keep the other two coordinates restricted to the Solar neighbourhood volume, i.e. $|R-R_0|<1$~kpc, $\phi-\phi_\odot<\pi/12$ and $|z|<2$~kpc. In all three figures, the top (bottom) row shows the data (simulation). The left-hand columns show the column-normalised density (such that the total count in each $R$, $\phi$ or $|z|$ bin is equal), and the right-hand columns show the antisymmetric components, defined similarly to $n_\mathrm{antisym}$.

The most interesting of these is the radial dependence (Fig.~\ref{fig:R_vR}). As shown by \citet{dillamore2024}, the resonances at low $|L_z|$ form a series of nested `chevrons'. The dashed coloured lines in the bottom-right panel of Fig.~\ref{fig:R_vR} indicate the principal resonances (CR, OLR, 1:1 and 3:2), which lie along the chevron-shaped overdensities. We describe the procedure for calculating these tracks in Section~\ref{section:fitting_model}. The overdensities are asymmetric, being only visible at $v_R<0$ for $R\lesssim15$~kpc, and only at $v_R>0$ for $R\gtrsim15$~kpc. This structure was explained and predicted analytically by our model in \citet{dillamore2024}, and is due to the shapes of the stable resonant orbits. The orbits shown in Fig.~\ref{fig:separatrices} can only be followed in one direction; the branches of the low-$L_z$ OLR (orange) and 1:1 (green) orbits that pass closest to the Sun are inward-moving, while those further out (in the Galactic anticentre direction) are outward-moving. The sign of the $v_R$ asymmetry therefore flips at the transition between these two branches, resulting in the pattern seen in the simulation in Fig.~\ref{fig:R_vR}.

The Iphicles and Beret moving groups can likewise be tracked across several kpc in $R$ in the top row of Fig.~\ref{fig:R_vR}. These features were previously discovered in this space by \citet{belokurov_chevrons}, who numbered them Chevrons 1 and 3 respectively, and suggested that they were formed by the phase-mixing of the \textit{Gaia} Sausage-Enceladus debris. However, the asymmetry of the features is inconsistent with an ancient ($>8$~Gyr) merger event \citep{donlon2023}, whereas this is a natural prediction of the resonant scenario proposed by \citet{dillamore2024}. This is further supported by the facts that Iphicles and the Beret align with the predicted locations of the CR and OLR at reasonable pattern speeds, and that they are seen at a wide range of metallicities (see Fig.~\ref{fig:data_metallicity}). In Section~\ref{section:fitting} we use the tracks of Iphicles and the Beret across a range of radii to simultaneously fit the potential of the Milky Way and the pattern speed of the bar.

Fig.~\ref{fig:phi_vR} shows $v_R$ as a function of azimuth $\phi$. The zero-points on the $x$-axes mark the position of the Sun $\phi_\odot$, and the angle of the bar's major axis \citep[$30^\circ$;][]{wegg2015} is shown with black dashed lines in the simulation panels. The CR and OLR features can be tracked across all azimuths in the simulation. However, the asymmetry changes sign every $\approx90^\circ$. The reason for this can again be seen from the shapes of the orbits in Fig.~\ref{fig:separatrices}. In both the low-$L_z$ CR and OLR cases, a branch of the orbit with the opposite sign of $v_R$ is encountered whenever the major or minor axis of the bar is crossed. For a steadily rotating bar, the asymmetry should therefore change sign at its principal axes. Fig.~\ref{fig:phi_vR} shows that this is not quite true for a decelerating bar; the flip occurs at a value of $\phi$ slightly larger than the major axis angle. This is because of the Euler force acting due to the deceleration of the rotating frame, causing the orbits to be twisted relative to the bar \citep[e.g., see Fig.~16 of ][]{chiba2021}. The Iphicles and Beret asymmetries can also be traced by eye over a small range of $\phi$, though the spatial limits of the \textit{Gaia} sample do not allow the behaviour beyond the major or minor axes to be seen. However, we may predict from the simulation that these streams have the opposite signs of $v_R$ at $\phi-\phi_\odot\gtrsim30^\circ$. If the angle of this sign change can be measured in the future, it may present a possible method to constrain the bar angle and deceleration rate.

Finally, Fig.~\ref{fig:z_vR} shows the dependence on distance from the Galactic plane, $|z|$. In both the simulation and the data, the CR (Iphicles) asymmetry can be traced as far as $|z|\sim4$~kpc, showing that this resonance extends to affect highly inclined orbits. The OLR feature is discernible as far as $|z|\sim3$~kpc in the simulation, while the Beret is clearly visible up to $|z|\approx1$~kpc. Both features tend towards smaller $|v_R|$ as $|z|$ increases, as expected if the resonances are at approximately constant energy.

\begin{figure}
  \centering
  \includegraphics[width=\columnwidth]{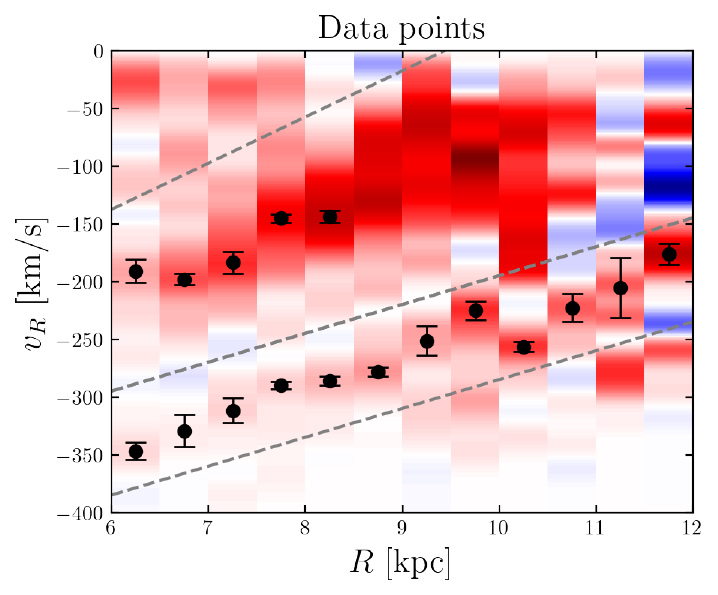}
  \caption{Data used for fitting orbits to the resonant streams in radial phase space $(R,v_R)$. The red and blue background shows kernel density estimations (KDEs) for the antisymmetric components $n_\mathrm{antisym}(v_R|R)$ of the distributions of stars in each $R$ bin. Red (blue) indicates where there is an excess (deficit) compared to the corresponding positive value of $v_R$. The grey dashed lines are manually selected boundaries around the Iphicles and Beret overdensities. The black data points are the most prominent peaks of $n_\mathrm{antisym}(v_R|R)$ in each of these two regions, with uncertainties calculated by bootstrap resampling.} 
   \label{fig:data_points}
\end{figure}

\section{Potential and pattern speed fitting}\label{section:fitting}
As Figs.~\ref{fig:separatrices} and \ref{fig:R_vR} demonstrate, the low-$L_z$ orbits trapped in the OLR explore a wide radial range of the Galaxy, from $R\approx0-18$~kpc. The ability to track these orbits in $(R,v_R)$ phase space via the $v_R$ asymmetry therefore offers a method to constrain the Galactic potential in these regions. This is analogous to fitting a potential to stellar stream observations \citep[e.g.][]{koposov2010}. The additional constraint that the orbits be trapped in resonances breaks degeneracies between the potential parameters, and allows us to simultaneously infer the bar's pattern speed $\Omegab$.

Our general approach is to fit resonant orbits to the $v_R$ asymmetries seen in radial phase space. The simulation shows that the exact planar resonant orbits in the azimuthally averaged potential align closely with these features (see Fig.~\ref{fig:R_vR}). We therefore fit CR and OLR orbits to the observed Iphicles and Beret asymmetries respectively. Our method and results are described in detail below, and we test the method on simulation data in Section~\ref{section:fitting_discussion}.

\subsection{Data points}\label{section:fitting_data}
We fit a model to data points derived from the antisymmetric components of the distributions in radial phase space (as in the right-hand column of Fig.~\ref{fig:R_vR}). With the same sample used to produce Fig.~\ref{fig:R_vR}, we divide the stars into 12 equally spaced bins in the interval $R\in\left[6,12\right]$~kpc. This is the range across which the Beret feature can be easily traced. To locate the peaks corresponding to each resonance, we divide the space into regions bounded by the lines
\begin{align}
    v_R/\left[\mathrm{km/s}\right]=\left\{\begin{array}{lr}
        40\\
        25\\
        25
        \end{array}\right\}\left(R-R_0\right)/\mathrm{kpc}
        -\left\{\begin{array}{lr}
        50\\
        240\\
        330
        \end{array}\right\}.
\end{align}
The two regions between each pair of neighbouring lines include the Iphicles and Beret features respectively. In each bin $i$ (centred at radius $R_i$) we fit a kernel density estimation (KDE) to the $v_R$ distribution of stars $n(v_R|R)$, using a Gaussian kernel of standard deviation $10$~km/s. We use this to calculate the antisymmetric distribution $n_\mathrm{antisym}(v_R|R)\equiv n(v_R|R)-n(-v_R|R)$. These KDEs are shown for $v_R<0$ in Fig.~\ref{fig:data_points}, with the grey dashed lines marking the boundaries described above. Using \textsc{scipy.signal.find\_peaks}, we find the peaks of $n_\mathrm{antisym}$ and their prominences. We take the locations of the most prominent peaks in each of the two bounded regions to be velocity measurements of the CR and OLR in that bin. Bootstrap resampling is used to repeat the KDE fitting and peak finding 500 times for each bin. For each resonance (indexed by $l$) we take the mean and standard deviation of the peak location across all samples as the measurement and its uncertainty, $v_{R,l,i}$ and $\sigma_{l,i}$.

This process is repeated in all 12 bins for the Beret ($l=1$), but in only 5 bins (those with $R_i\leq8.5$~kpc) for Iphicles ($l=0$). At larger radii the Iphicles stream becomes much broader as it approaches $v_R=0$, so there is no single well-defined peak that can be fitted reliably. We therefore exclude measurements of Iphicles at $R>8.5$~kpc. The peak measurements and their uncertainties are shown in black in Fig.~\ref{fig:data_points}.

Though we do not explicitly consider observational uncertainties in this calculation, scatter in the points resulting from observational errors will contribute to the variance of the peak locations. Distance uncertainties outside the immediate Solar neighbourhood are typically $\sim5\%$ in our selected data, resulting in velocity uncertainties from proper motions of $\sim0.05\vc\sim10$~km/s. This is comparable to the typical uncertainty we derive from bootstrap resampling, so our error estimates are reasonable. Also, much of the Galactocentric radial motion of stars in our sample is along the line of sight, so estimates of $v_R$ are dominated by \textit{Gaia} RVS measurements. These do not rely on distance measurements, and have higher precision \citep[typically $\sigma\lesssim6$~km/s;][]{gaia_rvs}.

\subsection{Model}\label{section:fitting_model}
Our potential model is based on the axisymmetric (i.e. azimuthally averaged) \citet{hunter2024} potential used throughout this paper. We keep the baryonic components $\Phi_\mathrm{baryon}(R,z)$ fixed, but allow the potential to vary by replacing the Einasto dark matter halo with a spherical Navarro-Frenk-White \citep[NFW;][]{NFW} potential described by two parameters,
\begin{equation}
    \Phi_\mathrm{NFW}(r)=-\frac{4\pi G\rho_0a^3}{r}\mathrm{log}\left(1+\frac{r}{a}\right).
\end{equation}
Instead of $\rho_0$ and $a$ we choose to parameterise the potential using the total circular velocity at the Sun's position,
\begin{equation}
    \vc\equiv v_\mathrm{c}(R_0)=\sqrt{R_0\left(\frac{\partial\Phi_\mathrm{baryon}}{\partial R}(R_0,0) + \frac{\partial\Phi_\mathrm{NFW}}{\partial r}(R_0)\right)},
\end{equation}
and the total mass enclosed within a radius of $r=20$~kpc,
\begin{equation}
    M_{20}\equiv M_\mathrm{baryon}(20\,\mathrm{kpc}) + M_\mathrm{NFW}(20\,\mathrm{kpc}),
\end{equation}
where $M_\mathrm{baryon}(r)$ and $M_\mathrm{NFW}(r)$ are the masses enclosed within radius $r$ of the baryonic and dark matter components respectively. $\vc$ and $M_{20}$ set the slopes of the potential at the Sun's position and near the apocentres of the eccentric OLR orbits respectively. They are therefore a suitable parameterisation for the regions explored by those orbits, and are more closely linked to observables than the NFW parameters $\rho_0$ and $a$. We choose to use an NFW rather than Einasto halo because it has fewer parameters to fit (two instead of three), while remaining a good model for Milky Way-like galaxies \citep[e.g.][]{mcmillan17}.

Our third parameter is the bar's pattern speed $\Omegab$, which determines a set of orbits in the potential $\Phi(R,z|\vc,M_{20})$ for each resonance. We restrict the orbits to the $z=0$ plane but leave the angular momentum $L_{z,l}$ as a free parameter to be marginalized over. This allows the angular momenta of the fitted orbits to freely vary over the $L_z$ range of the data. It ensures that the likelihood is not penalized by any incorrect assumptions about the $L_z$ distributions of the resonant features. In practice the choice of $L_z$ has only a weak effect on the track of the orbit (particularly the OLR), and we recover similar results when using a fixed value of $L_{z,l}=250$~kpc~km/s. 

Together, these four parameters define a single orbital track in $(R,v_R)$ space for the exact planar orbit at each resonance, $v_{R,l}(R|\vc,M_{20},\Omegab,L_{z,l})$. This satisfies the equation,
\begin{equation}\label{eq:vR_track}
    v_{R,l}^2(R)=2\left[E-\Phi(R,0)\right]-\frac{L_{z,l}^2}{R^2},
\end{equation}
where the energy $E$ of the orbit depends on $(\vc,M_{20},\Omegab,L_{z,l})$. We calculate the track by directly integrating a grid of orbits, calculating their distances from the resonance $-(\Omega_\phi-\Omegab)/\Omega_R-l/2$, and locating the gridpoints either side of the root of this curve. We then repeat with a finer grid between these points, and linearly interpolate to find the resonant orbit.

The bottom-right panel of Fig.~\ref{fig:R_vR} shows that there is a slight offset between the exact resonant orbits in the axisymmetric potential and the peaks of the $v_R$ asymmetry. Fig.~\ref{fig:separatrices} similarly shows that the low-$L_z$ exact resonant orbits are at slightly larger $|v_R|$ than the centres of the trapped regions. This is likely because of the difference in frequencies between orbits in the barred and axisymmetrised potentials. To account for this we include a linear $R$-dependent correction $m_lR+c_l$ in the model track used to fit the data, where $m_l$ and $c_l$ are nuisance parameters. The centroid of this model track is then
\begin{multline}
    v_{\mathrm{model},l}(R|\vc,M_{20},\Omegab,L_{z,l},m_l,c_l)=v_{R,l}(R|\vc,M_{20},\Omegab,L_{z,l})\\+m_lR+c_l
\end{multline}

Finally, we include one more nuisance parameter per resonance $\sigma_{\mathrm{model},l}$, which is simply a Gaussian width of the model in $v_R$. This smooths the model such that its probability distribution function at each value of $R$ is Gaussian instead of a delta function. Note that $\sigma_{\mathrm{model},l}$ does not describe the width of the resonance itself, but rather the spread of the peaks about the model track.

When fitting both the CR and OLR ($l\in\{0,1\}$), we therefore have a total of 11 parameters in the model. Two ($\vc$ and $M_{20}$) describe the potential; one ($\Omegab$) describes the bar's rotation; and eight ($L_{z,l}$, $m_l$, $c_l$, and $\sigma_{\mathrm{model},l}$) concern the tracks of the resonances.

\subsection{Likelihood}
The log-likelihood for the data point $(R_i,v_{R,l,i})$ with uncertainty $\sigma_{l,i}$ is
\begin{multline}
    \mathrm{log}\,\mathcal{L}_{l,i}=-\frac{1}{2}\mathrm{log}\left(2\pi(\sigma_{l,i}^2+\sigma_{\mathrm{model},l}^2)\right)\\-\frac{1}{2}\frac{\left(v_{R,l,i}-v_{\mathrm{model},l}(R_i)\right)^2}{\sigma_{l,i}^2+\sigma_{\mathrm{model},l}^2}.
\end{multline}
This is summed over all data points $i$ and the two resonances $l$ to give the total log-likelihood,
\begin{equation}
    \mathrm{log}\,\mathcal{L}=\sum_{l=0}^1\sum_{i=1}^{N_l}\mathrm{log}\,\mathcal{L}_{l,i},
\end{equation}
where $N_l\in\{5,12\}$ is the number of data points for the resonance $l$.

\begin{table}
\caption{Priors used for the MCMC fits. Units of $m_l$ and $c_l$ are km/s/kpc and km/s respectively. Their means and covariance matrices $\boldsymbol{\mu}$ and $\boldsymbol{\Sigma}$ are derived by calibration with the simulation.}
\begin{center}
\begin{tabular}{ c|c|c }
 \hline
 Parameter & Prior & Range \\
 \hline
 $\vc$ & Normal & $(229\pm6)$ km/s \\ 
 $M_{20}$ & Uniform & $(1,5)\times10^{11}M_\odot$ \\
 $\Omegab$ & Uniform & $(10,100)$ km/s/kpc \\
 $L_{z,l}$ & Uniform & $(0,500)$ kpc~km/s \\
 $(m_0,c_0)$ & Normal & $\boldsymbol{\mu}=(-2.01~\mathrm{kpc}^{-1},34.2)$~km/s,\\
 &&$\boldsymbol{\Sigma}=\begin{pmatrix}
49.6~\mathrm{kpc}^{-2} & -346~\mathrm{kpc}^{-1} \\
-346~\mathrm{kpc}^{-1} & 2440 
\end{pmatrix}$~(km/s)$^2$ \\[4mm]
 $(m_1,c_1)$ & Normal & $\boldsymbol{\mu}=(-1.32~\mathrm{kpc}^{-1},21.9)$~km/s,\\ &&$\boldsymbol{\Sigma}=\begin{pmatrix}
0.197~\mathrm{kpc}^{-2} & -1.61~\mathrm{kpc}^{-1} \\
-1.61~\mathrm{kpc}^{-1} & 13.6 
\end{pmatrix}$~(km/s)$^2$ \\[4mm]
 $\sigma_{\mathrm{model},l}$ & Uniform & $(0,100)$ km/s \\
 \hline
\end{tabular}
\label{tab:priors}
\end{center}
\end{table}

\begin{figure*}
  \centering
  \includegraphics[width=\textwidth]{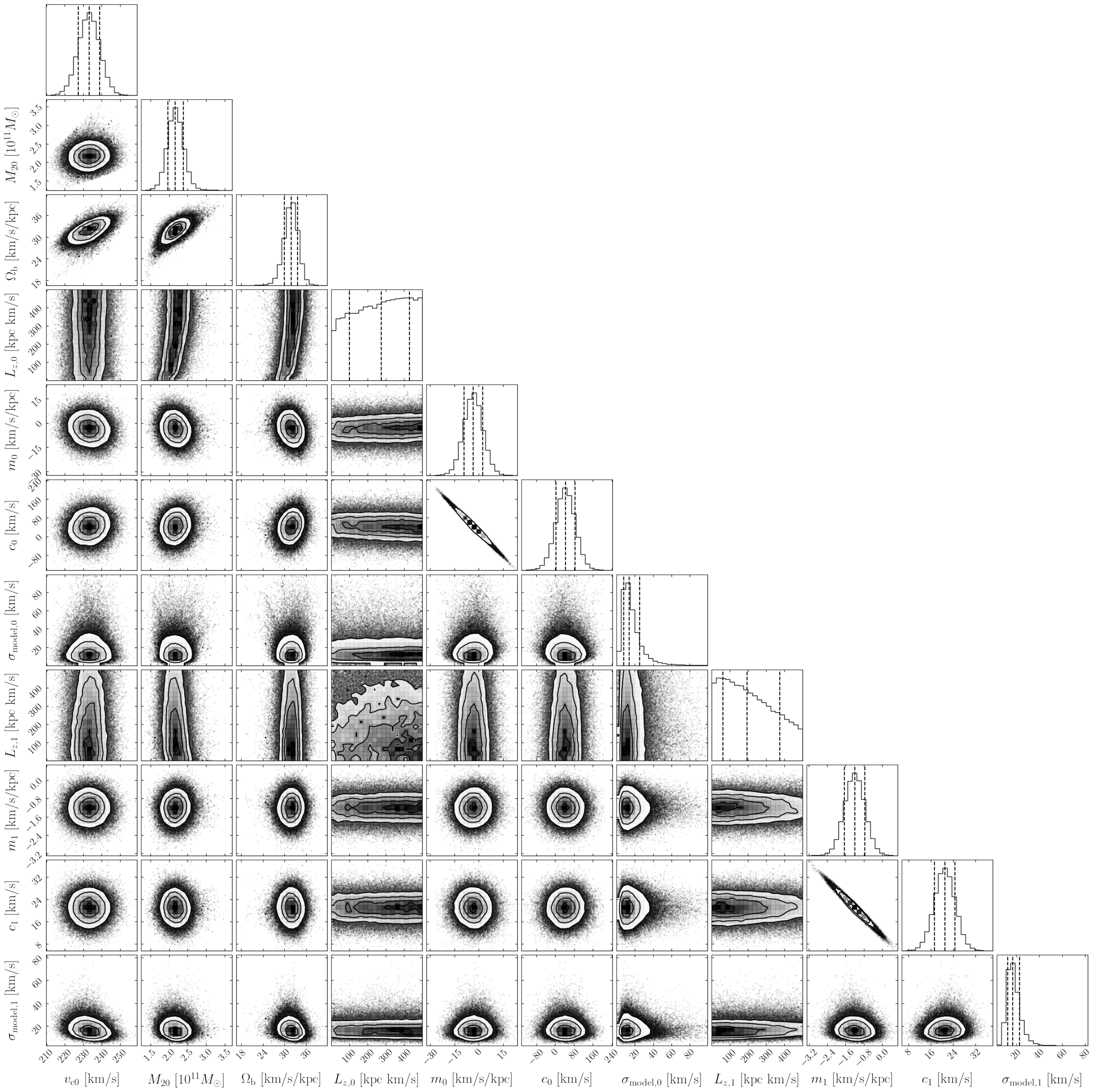}
  \caption{Posterior distributions of the fit to the data. The black dashed lines in the diagonal panels indicate the medians and 16th and 84th percentiles. All parameters are well-constrained except for $L_{z,0}$ and $L_{z,1}$, whose distributions are bounded by the $L_z$ range of the data.} 
   \label{fig:corner}
\end{figure*}

\begin{figure}
  \centering
  \includegraphics[width=\columnwidth]{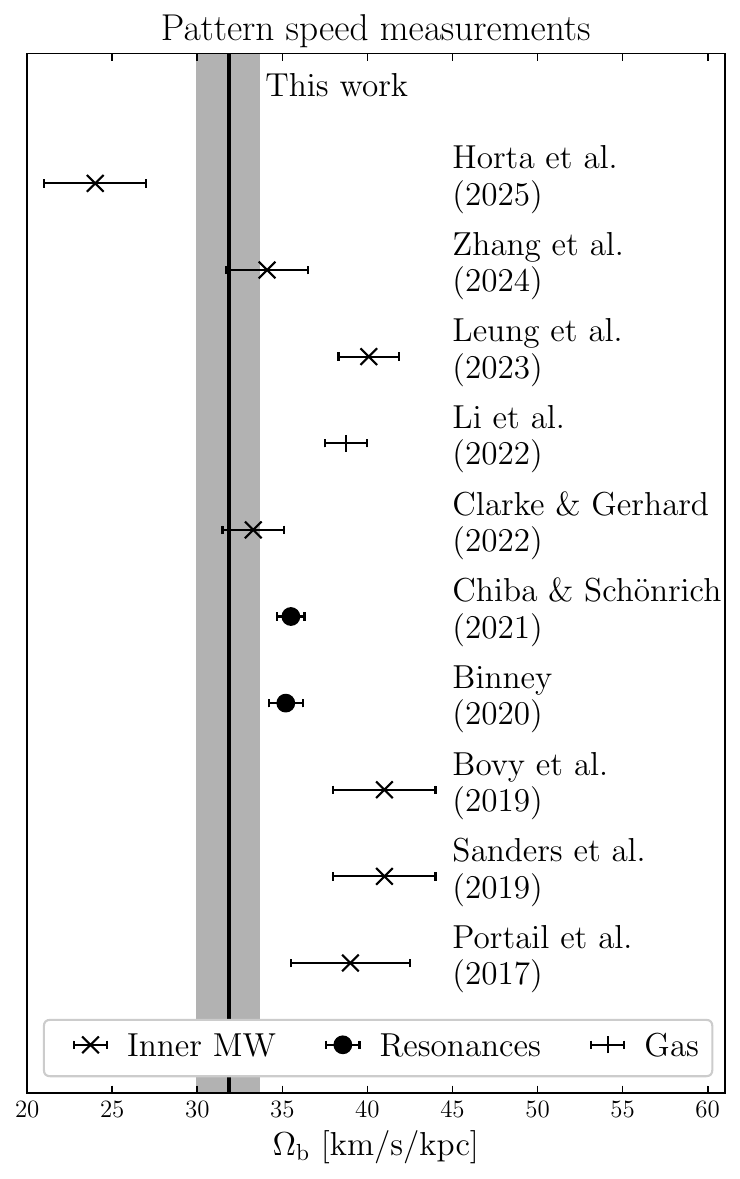}
  \caption{Comparison of our result for the pattern speed $\Omegab$ with several recent works. The marker type indicates whether the method employed direct observations of the inner Milky Way ($\times$), modelling of resonant features ($\bullet$), or gas measurements ($+$). From oldest to newest, they are \citet{Po17}, \citet{Sa19}, \citet{bovy2019}, \citet{binney2020}, \citet{chiba2021_treering}, \citet{Cl22}, \citet{li2022}, \citet{leung2023}, \citet{zhang24} and \citet{horta2024}.  With the exception of the latter, our favoured pattern speed is lower than these previous values, although it is consistent with both \citet{zhang24} and \citet{Cl22}.} 
   \label{fig:Omega_b_comparison}
\end{figure}

\begin{figure*}
  \centering
  \includegraphics[width=\textwidth]{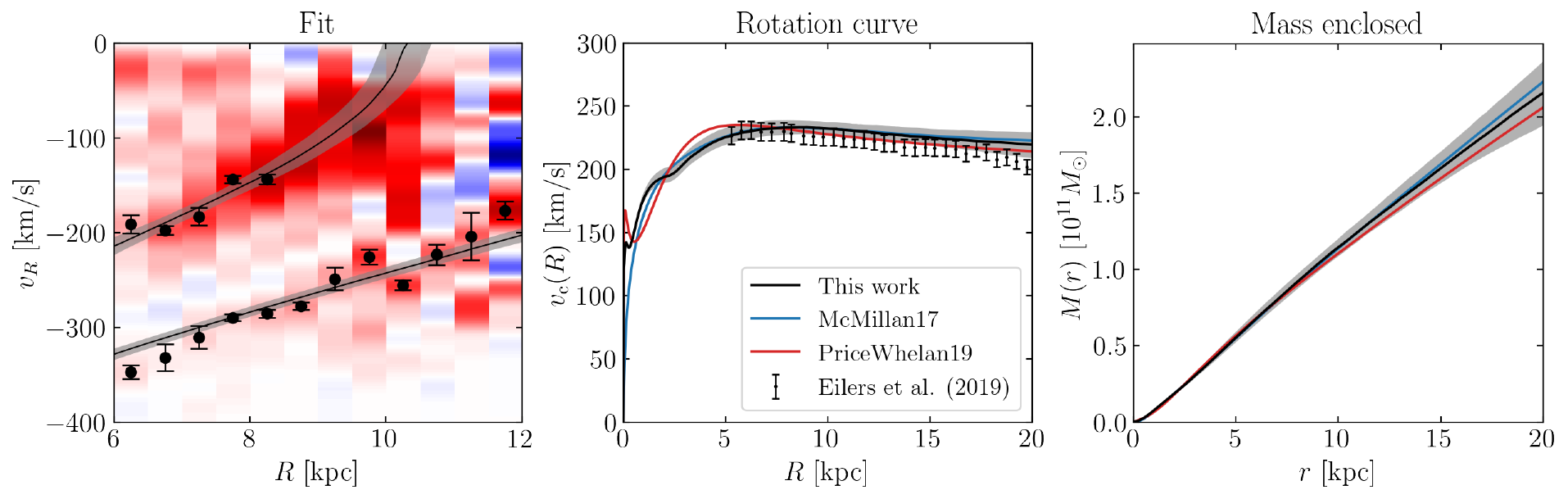}
  \caption{Best fitting models to the data using the prior on $\vc$ from \citet{eilers2019}. The black solid lines and grey bands indicate the medians and 16th-84th percentile ranges respectively. \textbf{Left-hand panel:} fit to the data points in radial phase space. We include the $m_lR+c_l$ correction, so these lines do not exactly correspond to the resonant orbits. The grey bands show the 16th-84th percentile ranges of the fits, not the model widths $\sigma_{\mathrm{model},l}$. \textbf{Middle panel:} Circular velocity curve $v_\mathrm{c}(R)$ of the fit compared to previous works. We show two widely used potential models, that of \citet{mcmillan17} and the \textsc{MilkyWayPotential} in \textsc{Gala} \citep{gala}. The scatter points with error bars are measurements of the rotation curve calculated by \citet{eilers2019} with the reported 3\% systematic uncertainties. \textbf{Right-hand panel:} The mass enclosed $M(r)$ within a sphere of radius $r$, compared to the two other potential models. There is excellent agreement between our fit and these other works, particularly the \citet{mcmillan17} potential.} 
   \label{fig:fit}
\end{figure*}

\subsection{Priors}
In order to provide independent constraints on the main parameters of our model, we use wide uniform priors for $M_{20}$ and $\Omegab$. Since the circular speed at the Sun's radius is better constrained by previous studies, we instead use a Gaussian prior for $\vc$. For our main model we take this from \citet{eilers2019}, derived from \textit{Gaia} DR2 and APOGEE \citep{apogee} data. Their value is $\vc=229$~km/s, with a systematic uncertainty of $\approx2-3\%$. We therefore use an uncertainty of 6~km/s as the standard deviation of the prior. Other methods have yielded different measurements of $\vc$; in Section~\ref{section:fitting_discussion} we discuss how our results are affected if a different prior for $\vc$ is used.

The angular momentum $L_{z,l}$ of the fitted orbits and model width $\sigma_{\mathrm{model},l}$ are both given uniform priors. $L_{z,l}$ is allowed to range between 0 and 500 kpc~km/s, the range used to produce the data points. The priors on $m_l$ and $c_l$ must however be restrictive, since they add a general linear function to the model track. We use normal priors with covariance between $m_l$ and $c_l$, obtained by calibrating with the simulation as follows. We take three snapshots of the simulation which roughly match the data, with pattern speeds of $\Omegab\in\{35.3,32.5,30.2\}$~km/s/kpc. In each snapshot we calculate the exact CR and OLR orbits across a uniform grid of $L_z$ between 0 and 500 kpc~km/s. Using a similar approach to that described in Section~\ref{section:fitting_data} (but without bootstrapping), we locate the antisymmetric peaks closest to these exact orbits. For each resonance, bin, snapshot and $L_z$ value we calculate the offset in $v_R$ between the peak and the resonant orbit, $\Delta v_{R,l,i}\equiv v_{R,l,i}-v_{R,l}(R_i)$. We then calculate the mean and standard deviation of all $\Delta v_R$ values in each bin for each resonance. These are treated as a measurement and its uncertainty of the offset between the exact resonant orbit and the peak of asymmetry. To produce the priors on $m_l$ and $c_l$, we use Markov chain Monte Carlo (MCMC) to fit a straight line $m_lR+c_l$ to the measurements $(R_i, \Delta v_{R,i})$. For this fit we use a Gaussian likelihood and wide uniform priors on $m_l$ and $c_l$. We calculate the means and covariance matrices of the posterior distributions, which we use to define the multivariate normal joint priors on $m_l$ and $c_l$ for fitting the observations. These are highly correlated due to the fact that $c_l$ is the value of the correction at $R=0$, not the average $R$ of the points. The priors for all variables in the main fit are given in Table~\ref{tab:priors}.

\subsection{MCMC setup}
Using Bayes' theorem we calculate the log-posterior from the total log-likelihood and log-priors, and run MCMC with \textsc{emcee} \citep{emcee}. We first maximise the log-likelihood with \textsc{scipy.optimize.minimize}, and initialise the walkers in a small volume around this solution. We use 50 walkers for a total of 15000 steps, and discard the first 2000 steps as a burn-in ($\sim8$ times the auto-correlation time). We present the results of this fit in Section~\ref{section:fitting_results}.

\subsection{Results}\label{section:fitting_results}
\begin{table}
\caption{Posteriors of the principal parameters of our model. The values shown are the medians, with the ranges indicating the 16th and 84th percentiles.}
\begin{center}
\begin{tabular}{ c|c|c }
 \hline
 Parameter & Posterior \\
 \hline
 $\vc$ [km/s] & $233.1_{-5.8}^{+5.7}$ \\[1.5mm]
 $M_{20}$ [$10^{11}M_\odot$]& $2.17_{-0.21}^{+0.21}$ \\[1.5mm]
 $\Omegab$ [km/s/kpc] & $31.9_{-1.9}^{+1.8}$ \\
 \hline
\end{tabular}
\label{tab:posteriors}
\end{center}
\end{table}
In Fig.~\ref{fig:corner} we show the posterior probability distributions for all 11 parameters in the model. The posteriors for $L_{z,l}$, $m_l$ and $c_l$ are dominated by the priors; the $L_{z,l}$ samples are spread across the full $0-500$~kpc~km/s range, and there are tight correlations between $m_l$ and $c_l$. There is also a positive correlation between $\vc$ and $\Omegab$. This is because higher circular speeds result in higher orbital frequencies, so require a higher $\Omegab$ for a given orbit to be in resonance. We check how the prior on $\vc$ affects the inferred pattern speed in Section~\ref{section:fitting_discussion}. The model widths $\sigma_{\mathrm{model},l}$ are small compared to the $v_{R,l,i}$ measurements, with the median less than 20~km/s for both resonances.

The principal parameters ($\vc$, $M_{20}$ and $\Omegab$) are all well-constrained, each with fractional uncertainties of less than $10\%$. In Table~\ref{tab:posteriors} we give the medians of the posteriors for these parameters, with uncertainties denoting the 16th and 84th percentiles. The posterior for the local circular speed $\vc$ is somewhat higher than the prior. The median is in fact almost equidistant between the estimates of \citet{eilers2019} (229~km/s) and \citet{gravity2019} (236.9~km/s), and both of these values are within our uncertainty. The result of our fit for $\vc$ is therefore fully consistent with previous results.

We estimate the mass enclosed within a radius of 20~kpc to be $M_{20}=(2.17\pm0.21)\times10^{11}M_\odot$. This is consistent with the smaller value of $1.91_{-0.17}^{+0.18}\times10^{11}M_\odot$ estimated from globular cluster kinematics by \citet{posti2019}. By contrast, \citet{malhan2019} measured a larger value of $(2.5\pm0.2)\times10^{11}M_\odot$ by modelling the GD-1 stream. Also using globular cluster motions, \citet{watkins2019} measures the mass enclosed within a radius of $r=21.1$~kpc, finding $M(21.1\,\mathrm{kpc})=2.1_{-0.3}^{+0.4}\times10^{11}M_\odot$. Extrapolating our fitted model gives a corresponding measurement of $M(21.1\,\mathrm{kpc})=2.28_{-2.3}^{+2.4}\times10^{11}M_\odot$, fully consistent with that of \citet{watkins2019}. \citet{kupper2015} used the Palomar 5 stream to measure the mass within $19$~kpc, obtaining $M(19\,\mathrm{kpc})=(2.1\pm0.4)\times10^{11}M_\odot$. Our corresponding estimate is $M(19\,\mathrm{kpc})=(2.07\pm0.19)\times10^{11}M_\odot$, again in close agreement. Our measurement of the total mass of the Milky Way within $r\sim20$~kpc is therefore generally consistent with previous results.

Our estimate for the pattern speed is $\Omegab=31.9_{-1.9}^{+1.8}$~km/s/kpc. In Fig.~\ref{fig:Omega_b_comparison} we compare this to other recent measurements. Our value is lower than most of these previous results, the only exception being that of \citet{horta2024} ($\Omegab=(24\pm3)$~km/s/kpc). Our result is consistent with both \citet{Cl22} and \citet{zhang24}, who measure $\Omegab=(33.3\pm1.8)$~km/s/kpc and $(34.1\pm2.4)$~km/s/kpc respectively. The earlier results favour slightly higher pattern speeds, in the range $35-41$~km/s/kpc. The value shown for \citet{Sa19} was derived using only observations of stars on the near side of the bar; the authors report that including stars on the far side reduces their measurement to $\Omegab=(31\pm1)$~km/s/kpc, which is fully consistent with our value. However, the results using this larger sample were inconsistent with observations of the Galactic centre, though small systematic changes in the proper motion measurements could change this. \citet{chiba2021_treering} derived their precise result of $\Omegab=(35.5\pm0.8)$~km/s/kpc using a metallicity gradient within the Hercules stream. However, they note that there are additional sources of systematic error not included in the quoted uncertainty. For example, they assume a simple logarithmic potential with a flat circular velocity curve at $v_\mathrm{c}=235$~km/s. They report that even a slightly inclined rotation curve can alter their measurement by $\sim1$~km/s/kpc, thus making it consistent with our result within the systematic uncertainty.

Data from \textit{Gaia} DR2 revealed that the Hercules stream is split into three branches at different values of $v_\phi$ \citep{gaia_dr2_disc}. \citet{binney2020} demonstrated that the highest-$v_\phi$ branches of the stream (at $v_\phi-\vc\approx-50$~km/s) coincide with the trapped CR region at $\Omegab\approx33$~km/s/kpc, while the branch at lower $v_\phi$ (at $v_\phi-\vc\approx-80$~km/s) would correspond to the CR at higher pattern speeds. The pattern speed inferred from the Hercules stream therefore depends on the interpretation of the different branches. In reality they may correspond to different resonances near the CR \citep[e.g.][]{asano2020}. In this case our results combined with \citet{binney2020} suggest that the branch at $v_\phi-\vc\approx-50$~km/s is most likely to correspond to the CR.

Fig.~\ref{fig:fit} shows the median and 16th-84th percentile range of the fit in three different spaces. The left-hand panel shows the fits in $(R,v_R)$ space compared to the antisymmetric distribution of stars and the data points $(R_{l,i},v_{R,l,i})$. The adjustment $m_lR+c_l$ is included in these tracks, but the uncertainty bands do not include the model width $\sigma_{\mathrm{model},l}$. Apart from the systematic offsets already accounted for, the orbits are a reasonable fit to the data points, passing along or close to the Iphicles and Beret streams across a wide range of radii. The largest discrepancy is that the observed points in the Beret would be better fit by a steeper line; i.e. an OLR orbit with larger $|\mathrm{d}v_R/\mathrm{d}R|$. We discuss this difference further in Section~\ref{section:fitting_discussion}.

The middle panel of Fig.~\ref{fig:fit} shows the rotation curve of our fitted model with associated uncertainties. For comparison we show the measurements by \citet{eilers2019}, where the error bars denote their estimated systematic uncertainty of $3\%$. We estimate a slightly higher circular speed at $r\sim20$~kpc than \citet{eilers2019}, though the results are just about consistent. We also show the rotation curves of two popular models for the Milky Way potential, namely the \citet{mcmillan17} potential and the \textsc{MilkyWayPotential} in \textsc{Gala} \citep{gala}. Similarly, the right-hand panel shows the mass enclosed $M(r)$ as a function of radius. There is excellent agreement between our fit and these models, particularly the \citet{mcmillan17} potential.

In summary, our fitted model for the potential is in excellent agreement with previous results, while we favour a slightly slower pattern speed $\Omegab$ than most recent works.

\begin{figure*}
  \centering
  \includegraphics[width=\textwidth]{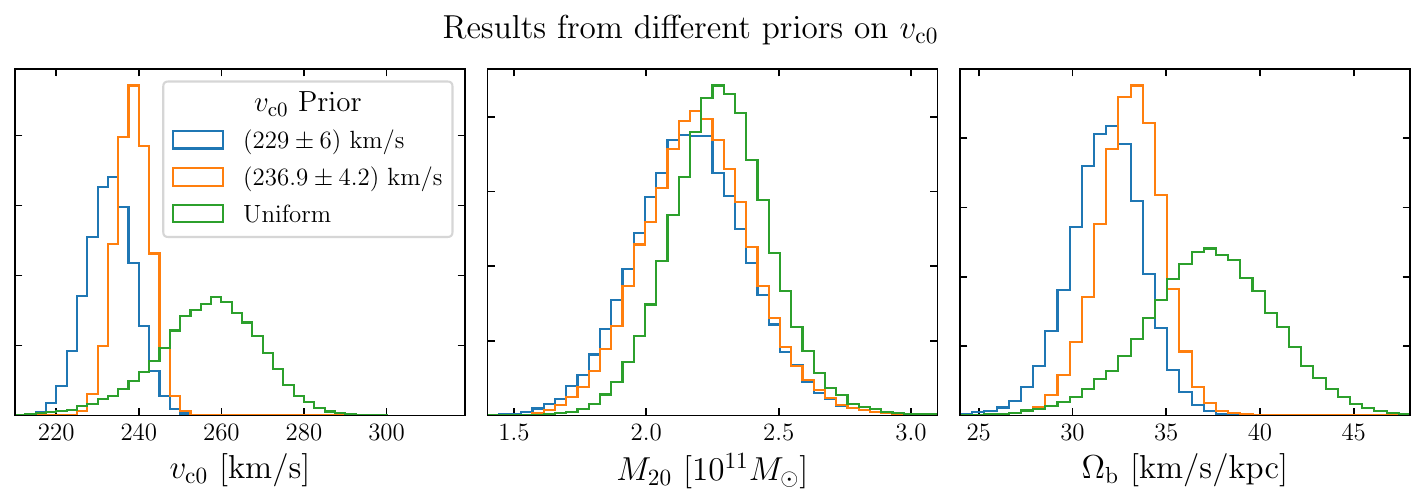}
  \caption{Posteriors on the three principal parameters on our model, from fits with different priors on $\vc$. Using a uniform prior results in the posteriors for both $\vc$ and $\Omegab$ being shifted to larger value, while $M_{20}$ is more robust.} 
   \label{fig:vc0_priors}
\end{figure*}

\begin{figure*}
  \centering
  \includegraphics[width=\textwidth]{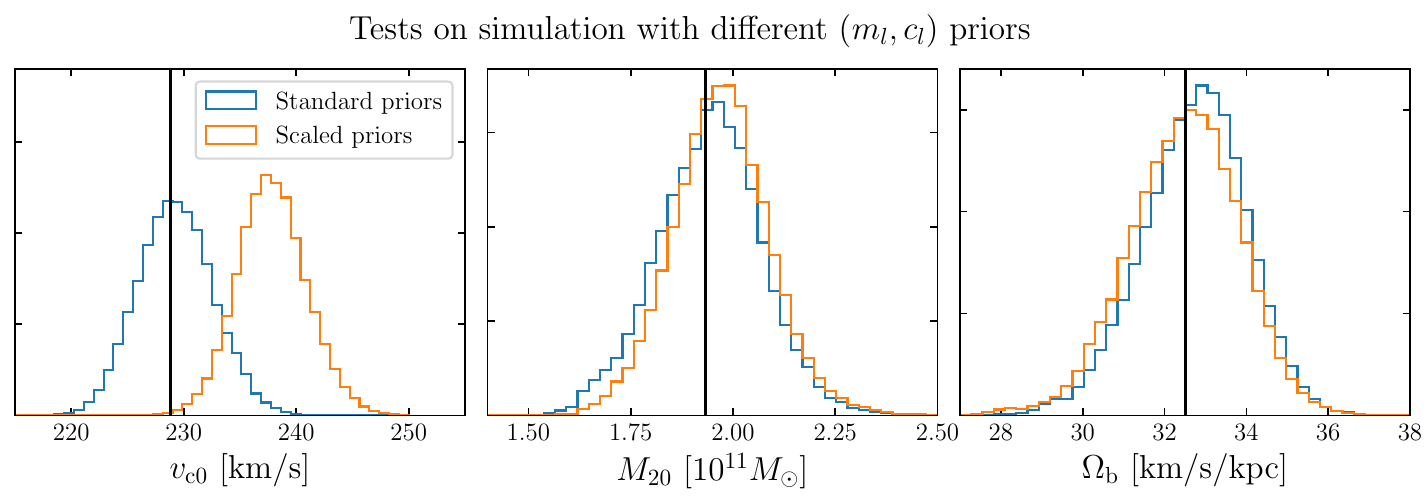}
  \caption{Posteriors on the three principal parameters on our model from fits on the simulation. In each panel the two histograms show the fits using different priors on $(m_l,c_l)$. The black vertical lines indicate the true values. With the $(m_l,c_l)$ calculated from the simulation, the fitting procedure recovers the correct values of $\vc$, $M_{20}$ and $\Omegab$. When these priors are manually altered by doubling the means of the Gaussians, $\vc$ is underestimated but $M_{20}$ and $\Omegab$ are still recovered accurately.} 
   \label{fig:sim_tests}
\end{figure*}

\subsection{Discussion}\label{section:fitting_discussion}
We now consider some of the main sources of systematic uncertainty in our fitting method, particularly from the informative priors used on $\vc$, $m_l$ and $c_l$. We also perform tests of the fit on the simulation to check that we can recover the correct parameters.
\subsubsection{Choice of $\vc$ prior}
In our main fit presented above, we use a normal prior taken from the measurement and uncertainty of \citet{eilers2019}, $\vc=(229\pm6)$~km/s. However, in principle we should be able to constrain $\vc$ from the value and slope of $v_R(R)$ at $R=R_0$. Differentiating equation~\eqref{eq:vR_track}, the gradient of an orbit in $(R,v_R)$ space satisfies,
\begin{align}
    v_{R,l}\frac{\mathrm{d}v_{R,l}}{\mathrm{d}R}&=-\frac{\mathrm{d}\Phi}{\mathrm{d}R}+\frac{L_{z,l}^2}{R^3}\\
    &=-\frac{v_\mathrm{c}^2}{R}+\frac{L_{z,l}^2}{R^3},
\end{align}
and the circular speed at $R=R_0$ is given by
\begin{equation}\label{eq:slope_vc0}
    \vc^2=\frac{L_{z,l}^2}{R_0^2}-R_0\,v_{R,l}(R_0)\frac{\mathrm{d}v_{R,l}}{\mathrm{d}R}(R_0).
\end{equation}
Hence in theory our measurements $(R_i,v_{R,l,i})$ should be sufficient to constrain $\vc$ using orbits with $L_{z,l}$ as free parameters, without even considering $M_{20}$ or the resonance condition. To test this we repeat the fit described above, but with a wide uniform prior on $\vc$. We show the posterior distributions for $\vc$, $M_{20}$ and $\Omegab$ in Fig.~\ref{fig:vc0_priors}.

The left-hand panel of Fig.~\ref{fig:vc0_priors} shows that when a uniform prior is used, the posterior distribution for $\vc$ is shifted to $\vc\approx(260\pm10)$~km/s. This is inconsistent with previous measurements \citep[see e.g.][]{bland-hawthorn2016}, and very likely results from the steep track of the Beret data points  $(R_i, v_{R,1,i})$ previously mentioned. This leads to a larger inferred value of $\vc$ via equation~\eqref{eq:slope_vc0}. This discrepancy may indicate that the priors on $m_l$ and $c_l$ calculated from the simulation do not fully capture the behaviour in the Milky Way. In the next section we therefore test our method on the simulation with different priors on $m_l$ and $c_l$, in order to assess the robustness of our $M_{20}$ and $\Omegab$ measurements. We should note that chevrons formed by phase mixing merger debris (as opposed to resonances) do not exactly align with orbits in radial phase space, due to energy sorting as a function of phase \citep[e.g.][]{dong-paez2022,davies2023_ironing}. However, in that case energy \emph{decreases} with increasing radial angle, so at $v_R<0$ the chevrons would have shallower slopes than the orbits. This is the reverse of what we see in our fit.

The large posterior on $\vc$ in the uniform prior case also leads to a faster inferred pattern speed, $\Omegab\approx(37\pm3)$~km/s/kpc. This is because larger circular speeds decrease the orbital periods, so the bar must rotate faster to maintain the resonance. However, the posterior on $M_{20}$ is more robust, only shifting by $\sim6\%$ to $M_{20}\approx(2.3\pm0.2)\times10^{11}M_\odot$. This overlaps strongly with the posterior from the fit with the \citet{eilers2019} prior. This is partly due to the relatively large fractional uncertainty on $M_{20}$, but also implies that the track of the OLR orbit strongly constrains the mass enclosed at large radii, with only weak dependence on the pattern speed and local circular speed.

There is some variation between recent measurements of $\vc$. \citet{gravity2019} report a value of $\vc=(236.9\pm4.2)$~km/s, based on their measurement of $R_0$, the proper motion of Sgr A* from \citet{reid2004}, and the velocity of the Sun relative to the local standard of rest (LSR) from \citet{bland-hawthorn2016}. This derivation assumes that the LSR is moving on a circular orbit at speed $\vc$, which may not be true given the non-axisymmetry of the Galaxy. To estimate the systematic uncertainty resulting from our choice of prior, we again repeat the fit with this normal prior, $\vc=(236.9\pm4.2)$~km/s. These results are also shown in Fig.~\ref{fig:vc0_priors}.

The effect of using the $\vc=(236.9\pm4.2)$~km/s is much milder, though as expected this also shifts the $\vc$ and $\Omegab$ posteriors to slightly larger values. In this case they are $\vc=(239\pm4)$~km/s and $\Omegab=33.1_{-1.7}^{+1.6}$~km/s/kpc. This would bring the pattern speed more in line with previous results (see Fig.~\ref{fig:Omega_b_comparison}). The posterior for $M_{20}$ is only slightly changed, to $M_{20}=(2.19\pm0.20)\times10^{11}M_\odot$. We can use the shifts in the posteriors to roughly estimate the systematic uncertainties resulting from our choice of $\vc$ prior: $\sim1$~km/s/kpc for $\Omegab$, and $\sim0.02\times10^{11}M_\odot$ for $M_{20}$. These are both smaller than our quoted uncertainties, so our results are reasonably robust to the choice of prior.

\subsubsection{Tests on the simulation}
We now test our fitting procedure on a snapshot of the simulation, to check that it can recover the correct parameters. The simulation was run using the potential model from \citet{hunter2024}, which has a dark matter halo with an Einasto profile \citep{einasto1965}. We can therefore also check whether our assumption of an NFW profile in the model biases the results. We first run the fit using an identical method to that used for the data, with the same priors. We use the fiducial snapshot of the simulation, with $\Omegab=32.5$~km/s/kpc. The posteriors on $\vc$, $M_{20}$ and $\Omegab$ are shown in Fig.~\ref{fig:sim_tests}.

The priors on $m_l$ and $c_l$ in the above fit were derived from the simulation we are testing on. Inherent features of the simulation such as the initial distribution function and bar time dependence are likely to influence these priors. Hence we must also test the sensitivity of the fit to these, and whether we can recover the correct values of $M_{20}$ and $\Omegab$ with incorrect priors on $m_l$ and $c_l$. We therefore repeat the fit on the simulation, but with the means of the $m_l$ and $c_l$ priors manually doubled. The posteriors from this fit are also shown in Fig.~\ref{fig:sim_tests}.

With the correct priors on $(m_l,c_l)$, all three parameters are recovered with excellent accuracy. This demonstrates that with correct priors on $m_l$ and $c_l$, our method is able to correctly infer $\vc$, $M_{20}$ and $\Omegab$, in spite of using a different model for the halo. Even when the scaled priors are used, the medians of the $M_{20}$ and $\Omegab$ posteriors are very close to the true values. The only incorrectly inferred parameter is $\vc$, which is overestimated by $\sim10$~km/s. This is because increasing the offsets shifts the fitted orbits to higher $|v_R|$ and $|\mathrm{d}v_R/\mathrm{d}R|$, which increases the inferred $\vc$ via equation~\eqref{eq:slope_vc0}. While this would increase the orbital frequencies, the fitted orbits are also shifted to greater apocentres and hence lower frequencies. These opposite effects are likely why the inferred $\Omegab$ is more robust than $\vc$. This test provides assurance that our estimates of $M_{20}$ and $\Omegab$ are not particularly sensitive to the $(m_l,c_l)$ prior used.

\subsubsection{Position and velocity of the Sun}
Throughout this paper we have used a fixed value for the distance to the Galactic centre, $R_0=8.178$~kpc \citep{gravity2019}, following \citet{hunter2024}. A more recent measurement is $R_0=(8.277\pm0.028)$~kpc \citep{gravity2021}, suggesting that the systematic uncertainty in the Sun's position is $\sim0.1$~kpc, or $\sim1\%$. This is somewhat smaller than both the fractional uncertainty in our prior for $\vc$, and the typical fractional uncertainty in the data points used in the fit. The total uncertainties in the Sun's velocity quoted by \citet{schonrich2010} are similarly small. Our fixed assumed values for these quantities are therefore unlikely to significantly contribute to systematic errors in our results.

\section{Conclusions}\label{section:conclusions}

In this paper we have investigated the resonant structure in velocity space of the stellar halo in the Solar neighbourhood. This has allowed us to constrain both the Milky Way's potential and the current pattern speed of the bar. Our principal conclusions are summarised below.

\begin{enumerate}[label=\textbf{(\roman*)}]
    \item At low angular momentum (small $|v_\phi|$), there are regions of velocity space in which orbits are trapped in resonances with the bar. Near the location of the Sun these are inward-moving (with $v_R<0$).

    \item By calculating the component of $(v_R, v_\phi)$ histograms which is antisymmetric in $v_R$ (i.e. $n_\mathrm{antisym}(v_R,v_\phi)\equiv n(v_R,v_\phi)-n(-v_R,v_\phi)$), we have shown that the observed velocity space near the Sun from \textit{Gaia} has multiple $v_R$-asymmetric features. In addition to the well-known disc moving groups (such as the Hercules and Sirius streams, and the `Horn' and the `Hat'), we reveal multiple asymmetries at low $|v_\phi|$. These are in the region of velocity space occupied by \textit{Gaia} Sausage-Enceladus. 
    
    \item The two most prominent asymmetric features peak at $(v_R,v_\phi)\approx(-150,50)$~km/s and $(-300,25)$~km/s, which we refer to as \textit{Iphicles} and the \textit{Beret} respectively (due to their respective associations with the Hercules stream and the Hat). These features closely correspond to the regions of velocity space trapped by the low-angular momentum corotation and outer Lindblad resonances respectively. Analogues of these two moving groups are produced naturally in our simulation by trapping in these resonances. The observed asymmetries persist across a wide range of metallicities, even down to [Fe/H]~$<-2$ \citep[as previously seen by][]{zhang2023_vmp}. We therefore conclude that these asymmetries are very likely dynamical streams produced by trapping halo stars in the bar's resonances. Nevertheless, most of the stars populating the streams are likely to have originated in the GSE, due to its dominance at low $L_z$.

    \item We trace the Iphicles and Beret asymmetries across a range of Galactocentric coordinates $R$, $\phi$ and $z$. In radial phase space $(R,v_R)$ the asymmetries manifest as the `chevrons' first discovered by \citet{belokurov_chevrons}. The Iphicles and Beret streams were there referred to as Chevrons 1 and 3 respectively. The resonant structure of $(R,v_R)$ space was previously investigated in detail by \citet{dillamore2024}, where we showed that the observed overdensities at $v_R<0$ are a natural consequence of trapping eccentric orbits in resonances. The streams can be traced spatially between $R\approx6-10$~kpc, across $\sim50^\circ$ azimuthally, and up to $|z|\sim4$~kpc from the Galactic plane. This is consistent with the predictions from the simulation.

    \item We fit resonant orbits to the tracks of the asymmetries in $(R,v_R)$ space to simultaneously constrain the mass of the Milky Way within 20~kpc $M_{20}$, as well as the bar's pattern speed $\Omegab$. We estimate the mass to be $M_{20}=(2.17\pm0.21)\times10^{11}M_\odot$, in excellent agreement with previous works \citep[e.g.][]{kupper2015,mcmillan17,watkins2019,posti2019,malhan2019}. Our value for the pattern speed is $\Omegab=31.9_{-1.9}^{+1.8}$~km/s/kpc, though this can be increased by $\sim1$~km/s/kpc depending on the prior used for the local circular speed $\vc$. This estimate is consistent with but slightly lower than most other recent constraints, which tend to favour values of $\Omegab\approx33-40$~km/s/kpc \citep[e.g.][]{Po17,Sa19,binney2020,chiba2021_treering,Cl22,zhang24}.
\end{enumerate}

This work is unique in providing quantitative constraints on both the bar and the dark matter halo to radii beyond the edge of the disc. Future \textit{Gaia} data releases and upcoming ground-based surveys such as WEAVE \citep{weave}, 4MOST \citep{4most} and LSST \citep{lsst} will allow halo substructure to be detected and measured to even greater distances. This will help to achieve the ultimate goal of a complete model of the Milky Way and its neighbourhood.

\section*{Acknowledgements}
We are grateful to the anonymous referee for a helpful report. We thank members of the Cambridge Streams group for useful comments and suggestions. AMD and JLS acknowledge support from the Royal Society (URF\textbackslash R1\textbackslash191555; URF\textbackslash R\textbackslash 241030). AMD and HZ thank the Science and Technology Facilities Council (STFC) for PhD studentships. 
VB acknowledges support from the Leverhulme Research Project Grant RPG-2021-205: "The Faint Universe Made Visible with Machine Learning".

\section*{Data Availability}
This paper uses publicly available \textit{Gaia} data. The code used in this project is available at \url{https://github.com/adllmr/resonances/tree/main}.



\bibliographystyle{mnras}
\bibliography{refs} 






\bsp	
\label{lastpage}
\end{document}